\documentclass[useAMS,usenatbib]{mnras}
\title[The low dark matter content of the lenticular galaxy NGC 3998]{The low dark matter content of the lenticular galaxy NGC 3998}
\author[N. F. Boardman et al.]{Nicholas Boardman$^{1}$\thanks{E-mail: nfb@st-andrews.ac.uk},
Anne-Marie Weijmans$^{1}$, Remco van den Bosch$^{2}$, Ling Zhu$^{2}$, \and Akin Yildirim$^{2}$,
Glenn van de Ven$^{2}$, Michele Cappellari$^{3}$, Tim de Zeeuw$^{4}$, \and Eric Emsellem$^{4}$,
Davor Krajnovi\'c$^{6}$, Thorsten Naab$^{7}$\\
$^{1}$School of Physics and Astronomy, University of St Andrews, KY16 9SS UK\\
$^{2}$Max Planck Institute for Astronomy, Königstuhl 17, D-69117 Heidelberg, Germany\\
$^{3}$Sub-department of Astrophysics, Department of Physics, University of Oxford, Denys Wilkinson Building, Keble Road, Oxford OX1 3RH\\
$^{4}$European Southern Observatory, Karl-Schwarzschild-Str. 2, 85748 Garching, Germany\\
$^{5}$Sterrewacht Leiden, Leiden University, Postbus 9513, 2300 RA, Leiden, The Netherlands\\
$^{6}$Leibniz-Institut f\"ur Astrophysik Potsdam (AIP), An der
Sternwarte 16, D-14482 Potsdam, Germany \\
$^{7}$Max-Planck-Institut f\"ur Astrophysik,
Karl-Schwarzschild-Str. 1, 85741 Garching, Germany\\}

\newcommand{\HI}{{\sc H\,i}}

\newcommand{\lagr}{\mathcal{L}}
\usepackage{anysize}
\usepackage{graphicx}
\usepackage{amsmath}
\usepackage[toc,page]{appendix}
\usepackage[]{hyperref}
\marginsize{2cm}{2cm}{2cm}{2cm}
\begin{document}

\date{Accepted 2016 May 13. Received 2016 May 13; in original form 2016 February 12}

\pagerange{\pageref{firstpage}--\pageref{lastpage}} \pubyear{2014}

\maketitle

\label{firstpage}

\begin{abstract}

We observed the lenticular galaxy NGC 3998 with the Mitchell Integral-Field Spectrograph and extracted line-of-sight velocity distributions out to 3 half-light radii. We constructed collisionless orbit models in order to constrain NGC 3998's dark and visible structure, using kinematics from both the Mitchell and SAURON instruments. We find NGC 3998 to be almost axisymmetric, seen nearly face on with a flattened intrinsic shape - i.e., a face-on fast-rotator. We find an I-band mass-to-light ratio of $4.7_{-0.45}^{+0.32}$ in good agreement with previous spectral fitting results for this galaxy. Our best-fit orbit model shows a both a bulge and a disc component, with a non-negligible counter-rotating component also evident. We find that relatively little dark matter is needed to model this galaxy, with an inferred dark mass fraction of just $(7.1^{+8.1}_{-7.1})\%$ within one half-light radius.

\end{abstract}


\section{Introduction}\label{intro}

Early-type galaxies (ETGs) are amongst the most highly-evolved structures in the Universe. Lying in the red sequence of galaxies, they are defined chiefly by their predominantly-old stellar populations. Imaging studies have repeatedly shown the massive ($M \geq 10^{11} M_\odot$)  ETG population to be smaller and more compact at $z \simeq 2$ than in the present day \citep{trujillo2006,cimatti2012,vandokkum2010}. Lower-mass galaxies appear to grow their mass in a far more uniform way, which suggests their growth to be dominated by in-situ star formation \citep{vandokkum2013}.

The ATLAS\textsuperscript{3D} survey \citep{cappellari2011} represents one of the first systematic studies of ETGs using integral-field spectroscopy. The survey includes two-dimensional spectroscopy with the SAURON instrument \citep{bacon2001}  for a volume-limited sample of 260 nearby ETGs, expanding upon the earlier SAURON survey\citep{dezeeuw2002}. SAURON observations have revealed a wide range of ETG kinematic structures. Massive ETGs mostly show little net rotation and are termed "slow rotators" (SRs), while less-massive ETGs are largely dominated by rotation and show evidence of stellar discs \citep{cappellari2007,emsellem2007,emsellem2011,kraj2013}; these are labelled "fast rotators" (FRs). Individual ETGs show a wide range of kinematic substructures, which include kinematic twists and kinematically-decoupled cores \citep{kraj2011}.

Analyses of the ATLAS\textsuperscript{3D} data have suggested that while FRs seem to smoothly connect to their later-type counterparts (including various and complex formation histories), SRs had a faster and more violent history \citep{cappellari2013b} SRs appear to form in an "inside-out" way, as in the "two-phase" model of ETG formation \citep{oser2010}: a central core forms at high redshift via dissipative processes \citep[e.g.][]{hoffman2010}, with late-stage growth then dominated by dry mergers and accretion  \citep[e.g.][]{naab2009,oser2010,oser2012,bai2014}. Compact high-redshift ETGs, meanwhile, are thought to have formed in a similar manner to nearby SRs \citep{barro2013,vandokkum2015}. This picture explains the strong correlations between the galaxy velocity dispersion $\sigma$ and other properties of local ETGs for galaxy masses below $\simeq 2\times 10^{11}M_\odot$ \citep{cappellari2013a}: $\sigma$ acts as a tracer for the bulge mass fraction, reflecting the growth of the bulge as a spiral becomes a fast-rotating ETG \citep[see, e.g.][for a review]{cappellari2016}. 

Several recent results have complicated the above picture, at least in regard to FRs. The concentration and stellar angular momentum of lenticular galaxies appear incompatible with them having passively evolved from spirals, implying that mergers are needed to reduce the angular momentum \citep{querejeta2015}. Major mergers are one possibility \citep{querejeta2015}, but simulations of repeated minor mergers have also produced realistic galaxies \citep{bournaud2007a}. The star-formation rate of star-forming galaxies in groups shows no evidence of being globally lower than for star forming galaxies in the field \citep{lin2014}, which implies quenching to be a rapid process; this further supports mergers as significant FR quenching mechanism. 

One way to further investigate the development of FRs is to study the kinematics of their outer regions. Dry accretion episodes,are expected to produce an excess of radial orbits beyond the central half-light radius (or effective radius, $R_e$) \citep{wu2014}, which is beyond SAURON's field of view (FOV) for most galaxies in the ATLAS\textsuperscript{3D} sample. Gas rich mergers and interactions, meanwhile, are expected to result in an excess of rotational orbits within a galaxy \citep{rottgers2014};

Kinematic studies of ETGs beyond 1 $R_e$ have yielded mixed results to date. \citep{arnold2014} present stellar kinematics for 22 nearby massive ETGs out to $~2-4 R_e$ and report a significant fraction of them to show abrupt drops in their angular momentum profiles beyond 1$R_e$; they argue that late dry accretion must have been important for these systems. \citep{raskutti2014} report much smoother angular momentum profiles for their ETGs out to $~2-5R_e$, though they note that the higher masses of their sample could be a potential confounding factor. Kinematic measurement for ETGs beyond 1$R_e$ have also been made using planetary nebular and/or globular cluster measurements \citep[e.g.][]{douglas2007,coccato2013,pota2013}, as well as using stellar measurements from long-slit spectra \citep[e.g.][]{kronawitter2000,khoperskov2014}.

\citet{schwarzschild1979} orbit-based dynamical modelling is an ideal tool for studying a galaxy's kinematic structure in further detail. Here, a galaxy is modelled as a superposition of collisionless stellar orbits, with no prior assumptions made on the nature of these orbits.  Orbit-based dynamical modelling allows various galaxy properties such as the stellar mass-to-light ratio and the galaxy viewing angle to be constrained, and it also enables the distribution of stellar orbits to be studied as a function of position.

The role of dark matter is also important to consider here. From the hierarchical cold dark matter paradigm \citep{blumenthal1984}, all ETGs should exist within massive dark matter halos.  Dynamical modelling of ATLAS\textsuperscript{3D} data has yielded a median dark matter fraction of 13\% within $1R_e$\citep{cappellari2013}; thus, dark matter must be considered when constructing wide-field dynamical models of ETGs. 

Orbit modelling has been used to investigate ETG dark halos in the past \citep[e.g.][]{weijmans2009,cappellari2013}, but such studies remain rare for complete two-dimensional stellar kinematics beyond the central effective radius. \citet{yildirim2015} model two compact ETGs out to 3-4 $R_e$, and find dark matter to be necessary to model one (MRK 1216) but not the other (NGC 1277). \citet{yildirim2016} model the compact ETG NGC 1286 out to 5$R_e$ and also find models with a massive dark halo to fit the galaxy best. 

The Mitchell Integral-Field Spectrograph \citep{hill2008}, formerly VIRUS-P, is particularly well-suited to investigating ETGs' structures beyond 1 $R_e$.  It features a $1.68 \times 1.68$ arcminute field of view, as well as large fibres (radius 2.08$''$) which reduce the need for spatial binning. 

In this paper, we focus on a single ETG: the lenticular galaxy NGC 3998, which we observed with the Mitchell Spectrograph over several nights. We extracted kinematics out to three $R_e$ up to the fourth Gauss-Hermite moment. We then used triaxial collisionless orbit models \citep{vandenbosch2008} to constrain the dark matter content, intrinsic shape, stellar mass-to-light ratio (M/L) and orbital structure of this galaxy. This paper serves to showcase our Mitchell Spectrograph data as well as a variety of techniques we are employing; we will present similar work on a larger set of galaxies in a future publication.

The structure of this paper is as follows. We discuss our observations and data reduction  in~\autoref{sample}. We describe the extraction of kinematic data in~\autoref{kin}, and we discuss the orbit-modelling method in~\autoref{mod}. We present the results of our modelling in~\autoref{res}. We discuss our findings in \autoref{disc} and then conclude in \autoref{sum}.

We assume a Hubble parameter of $h = 0.73$ throughout this work.

\section{Sample and data reduction}\label{sample}

\begin{table}
\begin{center}
\begin{tabular}{|c|c|}
\hline 
Type & S0 \\ 
\hline 
RA($^\circ$)(J2000) & 179.484039 \\ 
\hline 
DEC($^\circ$)(J2000) & 55.453564 \\ 
\hline 
Distance(Mpc) & 13.7 \\ 
\hline 
log($R_e$)($''$) & 1.30 \\ 
\hline
log($L_I$)($L_\odot$) & 10.15 \\ 
\hline
 $M_{bh}$ &  $8.1 \times 10^8 M_\odot$\\
\hline
\end{tabular} 
\end{center}
\caption{Summary of NGC 3998's properties. The black hole mass $M_{bh}$ and I-band luminosity $L_I$ are from \citet{walsh2012}. All other values are from \citet{cappellari2011} and references therein.}
\label{tab1}
\end{table}

NGC 3998 is a bright, nearby \citep[13.7 Mpc;][and references therein]{cappellari2011} lenticular galaxy with a powerful LINER nucleus in its centre \citep[e.g.][]{dressel2001}. The galaxy's brightness, combined with its simple morphology, makes it an ideal target for dynamical modelling \citep[e.g.][]{walsh2012}. NGC 3998 is part of the ATLAS\textsuperscript{3D} sample and has been detected in \HI\ \citep{serra2012}. We provide a brief summary of NGC 3998 in \autoref{tab1}.

\begin{figure}
\begin{center}
	\includegraphics[trim = 5cm 14cm 6cm 4cm,scale=0.5]{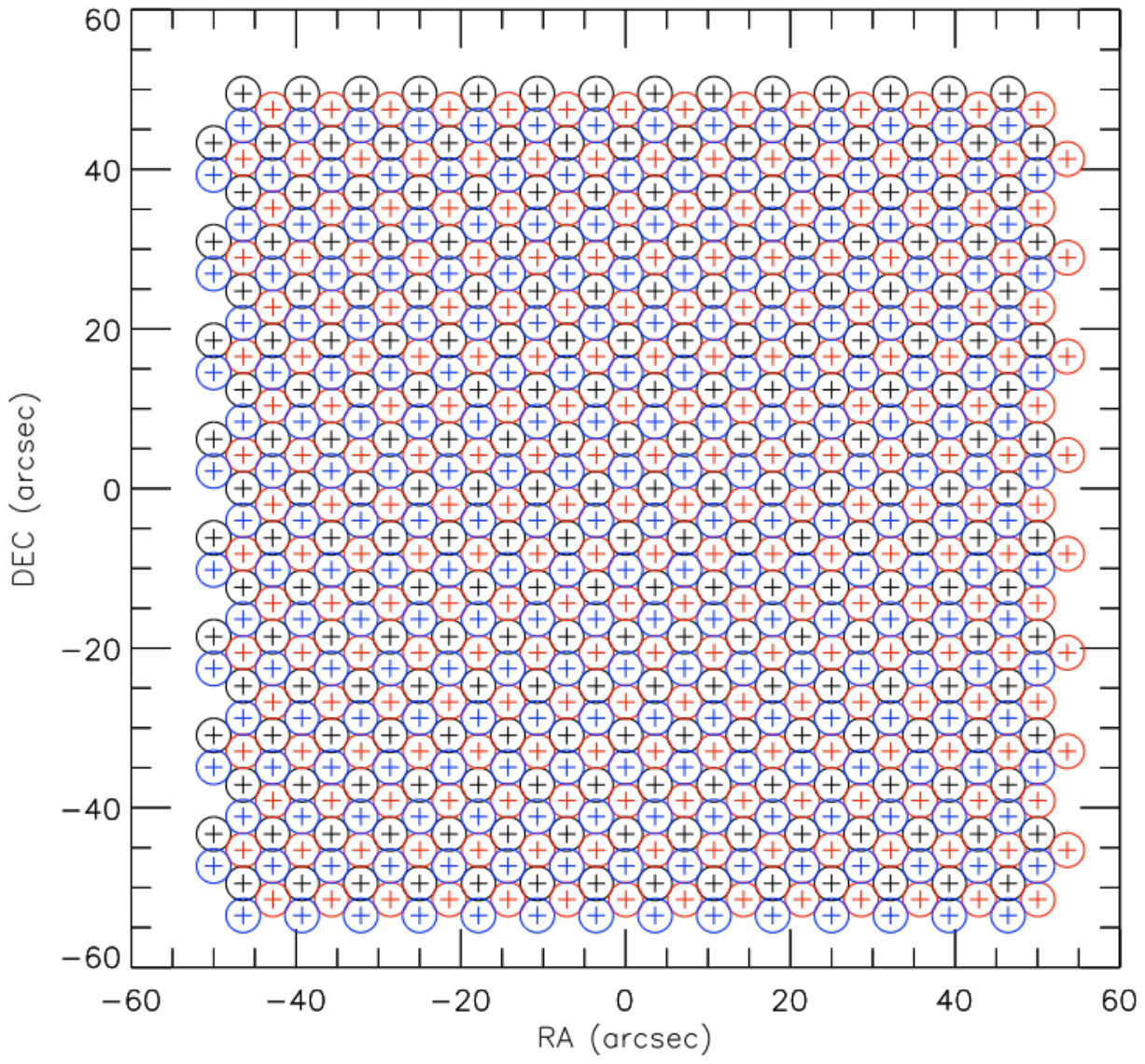}
	\caption{Mitchell Spectrograph first (black crosses), second (red crosses) and third (blue crosses) dither positions. The circles around each point represent the fibre cross sections. We achieve good sky coverage by combining all three dithers}
	\label{ditherfig}
	\end{center}
\end{figure}

We took 38 exposures of NGC 3998 with the Mitchell Spectrograph \citep{hill2008} in the high-resolution red setup (spectral range 4770\AA\ - 5430\AA), over nine nights between the 9th and 18th of March, 2010. Each exposure was taken for 1800s, producing a total time on source of 19 hours.   We observed at three dither positions in order to fully sample the field of view; this is illustrated in \autoref{ditherfig}. Since the spectrograph lacks dedicated sky fibres, we took sky exposures of 900s each so as to bracket pairs of science observations. We took bias frames, flat frames and arc frames at the beginning and end of each night. We selected either the dawn or dusk arcs and flats for each night to minimise the temperature differences between those and the observations.  We used Ne+Cd comparison lamps for all arc frames. 

We  carried out most of the data reduction using the VACCINE pipeline \citep{adams2011}. VACCINE is designed to avoid resampling of data, which can result in propagated errors, and uses techniques similar to those proposed by \citet{kelson2003}. It subtracts science frames for overscan and bias, calculates tracing and wavelength solutions for each fibre, calculates a corresponding sky frame for each science frame and collapses each fibre down into a 1D spectrum. Our reduction also included the LA-Cosmic algorithm of \citet{vandokkum2001}, which is designed to mask all frames of cosmic rays. 

We calculated the spectral resolution of the instrument using the acquired arc frames. We fitted Gaussians to the 5154.660 and 5400.56\AA\ emission lines across all nine nights, in order to obtain the spectral full-width at half maximum (FWHM) as a function of fibre position. We found the FWHM to vary smoothly as a function of fibre position, with values ranging from 1.26 to 1.63\AA\ . Our maximum FWHM corresponds to an intrinsic instrumental velocity dispersion of $\sigma = 41$km/s.

We constructed sky frames from sky exposures taken before and after a given galaxy frame. We used the method and code of \citet{blanc2013}, which allows for non-linear variations of the sky spectrum in both time and wavelength space, and then sky-subtracted all science frames. We combined all reduced science frames into a single spectral datacube, in which all spectra were interpolated onto a common linear scale. We then broadened all spectra in the cube to a common FWHM of 1.63\AA .

We masked three fibres located at the edges of the CCD chip during data reduction, in order to remove fibres not completely on the chip. These fibres are all located at the edges of our FOV, and so removing them does not have a big impact on our data. We also excluded two fibre positions from our analysis which we found to be contaminated by foreground objects.

\section{Kinematics}\label{kin}

In this section, we describe the extraction of NGC 3998's line-of-sight stellar kinematics up to the fourth Gauss-Hermite moment $(V,\sigma,h_3,h_4)$. In section 3.1, we provide an overview of our chosen method and present our results obtained from the Mitchell Spectrograph. In section 3.2, we report our re-extraction of SAURON kinematics with our chosen stellar library, for maximum consistency between the two instruments' results.

\subsection{Mitchell Kinematics}

To improve the signal-to-noise (S/N) at the galaxy's outermost regions, we binned spectra using the publicly available Voronoi Binning algorithm \citep{cappellari2003}. We chose a minimum S/N target of 40 per spectral resolution element to ensure that all four kinematic moments could be reliably extracted; this target was already satisfied for many of the spectra, resulting in relatively little binning.

We extracted stellar kinematics using the publicly available penalised pixel fitting (pPXF) software \citep{cappellari2004}. The pPXF routine fits an optimised template $G_{mod}(x)$ to a galaxy spectrum $G(x)$ directly in pixel space after logarithmically rebinning the spectrum in the wavelength direction, thereby recovering the LOSVD of that spectrum.  The model spectra take the form

\begin{equation}\label{ppxf2}
G_{mod}(x) = \sum_{k=1}^K w_k[\lagr (cx)*T_k](x) + \sum_{l=0}^L b_l\mathcal{P}_l(x)
\end{equation}

where $\lagr(cx)$ is the broadening function, $T_k$ a set of distinct stellar templates and $w_k$ the optimal weights of those templates, with $*$ describing convolution.  $\mathcal{P}_l(x)$ are Legendre polynomials of order $l$ and are used to adjust for low-frequency differences between model and data, with $b_l$ the corresponding weights.

The pPXF routine requires an input "bias parameter" which biases the recovered LOSVD towards a perfect gaussian when $h_3$ and $h_4$ become ill-determined, in order to prevent spurious solutions. We optimised this parameter using the simulation code included in the pPXF package, with the standard prescription that the deviation between input and output $h_3$ and $h_4$ should be less than rms/3 for values of $\sigma$ greater than three times the velocity scale. This led to an optimal bias of 0.2 for our target S/N of 40.

\begin{figure}
\begin{center}
	\includegraphics[trim = 5cm 0cm 5cm 4cm, scale=0.45]{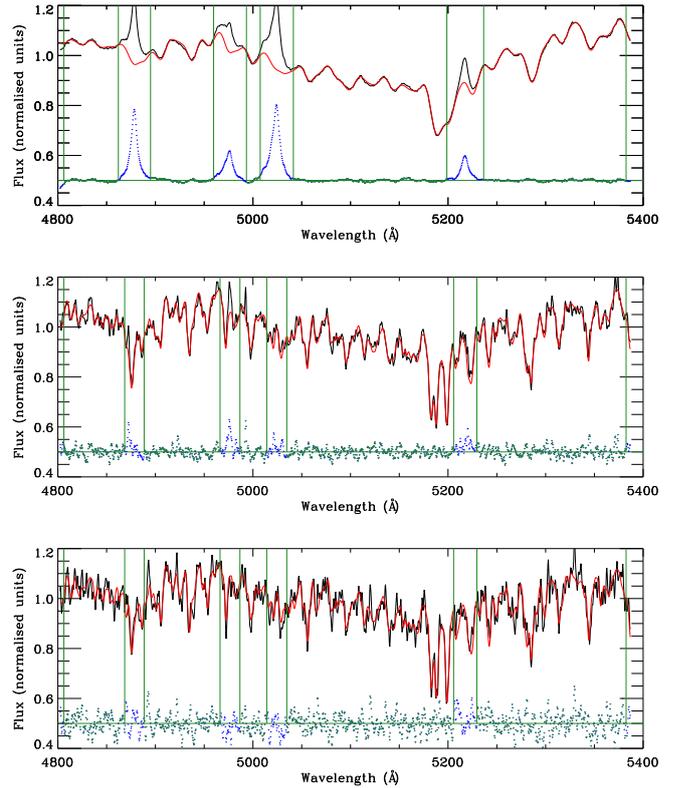}
	\caption{Example four-moment pPXF fits for the centremost fibre (top), a $2 R_e$ spectral bin (middle) and a $3 R_e$ spectral bin (bottom). Black lines show the normalised spectra and red lines the best fits. Vertical green lines indicate the positions of emission lines which we excluded from the fit. The blue line shows the residuals, with the horizontal green line marking the zero-point.}
	\label{residcheck}
	\end{center}
\end{figure}

For stellar templates, we used the medium-resolution (FHWM $= 0.51$\AA)  ELODIE library \citep{prugniel2001} of observed stars. We selected stars from the ELODIE library by performing a set of initial pPXF fits on elliptical annuli, which were constructed using ATLAS\textsuperscript{3D} ellipticity data \citep{kraj2011}; this resulted in a total of 31 stars being selected. We then ran pPXF over all binned spectra, with ionised gas emission regions (H$\beta$, [OIII] and [NI]) masked out. We allowed for ten additive Legendre polynomials when performing the fits. We determined measurement errors by adding Gaussian noise to the spectra and rerunning the fits for 100 iterations each, with the pPXF penalty set to zero as prescribed in section 3.4 of \citet{cappellari2004}. We present three example pPXF fits in \autoref{residcheck}.

\begin{figure}
\begin{center}
	\includegraphics[trim = 5cm 0cm 5cm 4cm, scale=0.45]{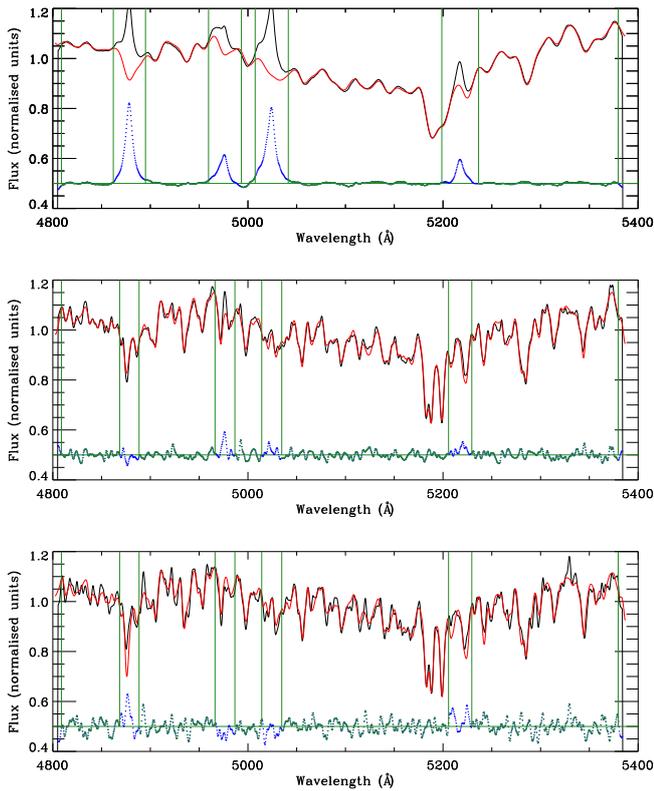}
	\caption{As in \autoref{residcheck}, but using MILES stars as templates instead of ELODIE The Mitchell spectra have been broadened to match the MILES resolution, as discussed in the text.}
	\label{residcheckmiles}
	\end{center}
\end{figure}

\begin{figure}
\begin{center}
	\includegraphics[trim = 2cm 13cm 0cm 2cm, scale=0.5]{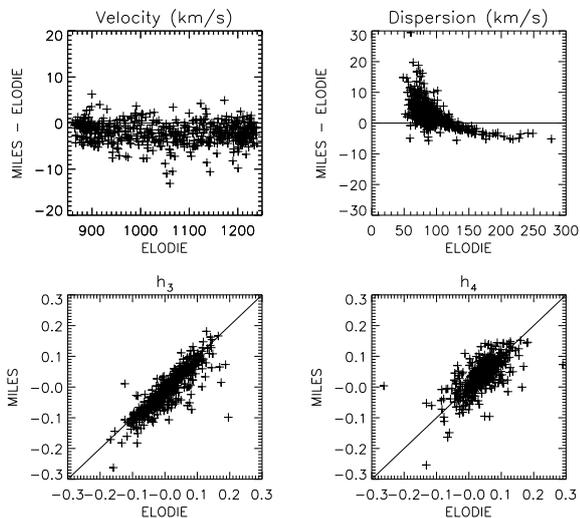}
	\caption{Comparison between kinematics calculated from Mitchell using ELODIE and MILES  templates. We find non-negligible differences in the results, suggesting the need to consider systematic errors in our work.}
	\label{templatecompare}
	\end{center}
\end{figure}

We are using deep spectroscopic observations that go well below the sky level; this means that our data will be dominated by systematics - such as imperfect sky substraction and template mismatch - and that Monte-Carlo simulations will underestimate the true errors on the kinematics \citep{arnold2014}. We therefore evaluated the systematic errors on our kinematics, by running pPXF for a second time, using stars from the MILES library of stars instead of ELODIE. The MILES library has an estimated resolution of $2.5\AA\ $ FWHM and so we broadened our spectra accordingly for this run. We selected a sample of MILES stars by fitting to elliptical annuli in the same manner as for the ELODIE stars, selecting 29 stars in total, and then we performed pPXF fits on all spectral bins. We show example fits with MILES in \autoref{residcheckmiles} and we compare the ELODIE and MILES kinematics in \autoref{templatecompare}.

We find non-negligible offsets between our two runs for all four kinematic moments, suggesting the need to include systematic errors in our analysis. We estimated systematic errors of 4 km/s and 5 km/s for our velocity and dispersion, and we estimated systematic errors of 0.03 for $h_3$ and $h_4$. We added these errors in quadrature to the initial error values. We found that a minority of kinematic values continued to exhibit large offsets relative to their errors once this process was completed; we masked the affected values from all subsequent analysis. 

\begin{figure*}
\begin{center}
	\includegraphics[trim = 0cm 9cm 0cm 0cm,scale=0.6]{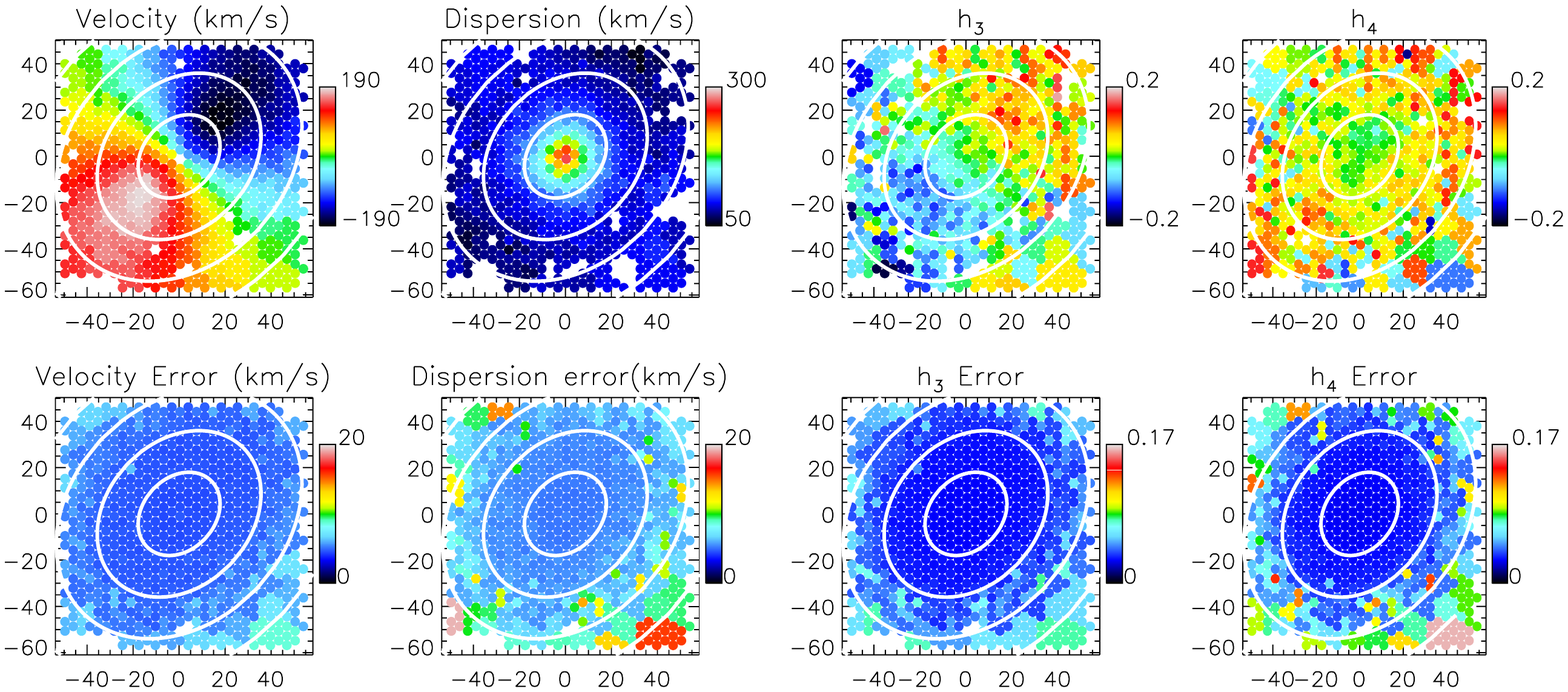}
	\caption{LOSVD and error maps measured for four moments from Mitchell Spectrograph data. The white contours are spaced in units of effective radii. We find evidence of ordered rotation through the FOV, with $h_3$ taking the opposite sign to the velocity. White regions show bins/fibres that were excluded, as discussed in the text.}
	\label{vd4mom}
	\end{center}
\end{figure*}

We present the resulting LOSVD maps in \autoref{vd4mom} along with the associated errors. We find evidence of ordered rotation throughout the FOV. \autoref{vd4mom} implies an anticorrelation between $V/\sigma$ and $h_3$, which may indicate a rotating axisymmetric disc component within a slower-rotating bulge \citep[e.g.][]{bender1994,naab2014}; this is expected for a fast-rotator like NGC 3998. Our kinematics extend beyond $3 R_e$, and thus cover a significantly wider FOV than previous IFU observations of this galaxy. We find a peak velocity of 190 km/s and a peak velocity dispersion of 277 km/s.

\subsection{SAURON Kinematics}

To test the accuracy of our kinematic extraction, we compared our velocity dispersion results to those of the ATLAS\textsuperscript{3D} survey \citep{cappellari2011} \footnote{Data available from http://purl.org/atlas3d} in the region where the datasets overlap. We mapped each of our fibres or spectral bins to the luminosity-weighted average of all enclosed SAURON pixels or bins, after realigning the datasets as explained in section 4.2. We show the results of this process in \autoref{losvdcompare1}. The ATLAS\textsuperscript{3D} velocity dispersions are noticeably higher at lower Mitchell dispersion values, but we note that all such offsets occur in regions where the ATLAS\textsuperscript{3D} data has been binned. 

\begin{figure}
\begin{center}
	\includegraphics[trim = 2cm 13cm 0cm 2cm, scale=0.5]{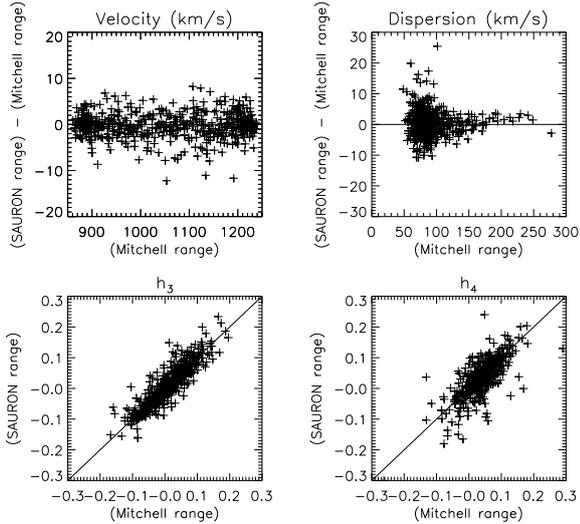}
	\caption{Comparison between kinematics calculated when fitting Mitchell spectra with ELODIE stars over the full available wavelength range vs kinematics calculated over the SAURON wavelength range. Black lines mark the 1-1 relation. We find that the wavelength range has little overall effect, with the scatter effectively captured by our derived uncertainties.}
	\label{saurangecheck}
	\end{center}
\end{figure}

\begin{figure*}
\begin{center}
	\includegraphics[trim = 0cm 13cm 2cm 7cm, scale=0.8]{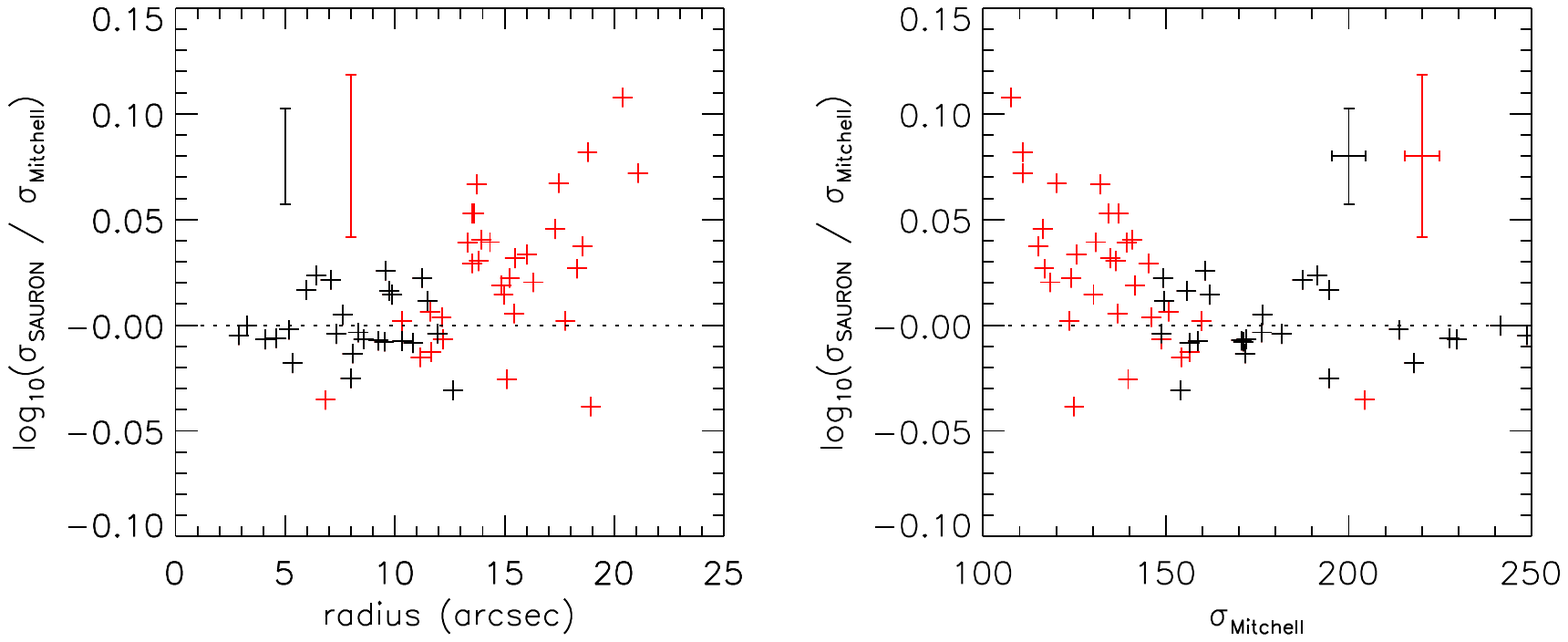}
	\caption{Left: ratio of ATLAS\textsuperscript{3D} to Mitchell Spectrograph velocity dispersions, plotted against radius (left) and the Mitchell velocity dispersion (right). The ATLAS\textsuperscript{3D} values are given as the flux-weighted averages of all values within a given Mitchell fibre. Black crosses on the left plot represent regions where no voronoi-binning took place, while red crosses indicate where the ATLAS\textsuperscript{3D} results were binned. We show representative error bars on both plots. We observe notable offsets at low Mitchell dispersions, where the ATLAS\textsuperscript{3D} spectra were binned.}
	\label{losvdcompare1}
	\end{center}
\end{figure*}

There are a number of caveats concerning this comparison. Since the releative offsets are largest in the binned ATLAS\textsuperscript{3D} regions, it is possible that the offsets are due to S/N differences between the datasets \citep[e.g.][]{arnold2014}. The velocity resolution of the datasets are also very different: the ATLAS\textsuperscript{3D}  data has an instrumental dispersion of 95km/s \citep{cappellari2011}, while the value for the finalised Mitchell spectra is 41km/s. This could further increase dispersion offsets when the dispersion approaches the ATLAS\textsuperscript{3D}  resolution. Differences in wavelength coverage are another consideration \citep{foster2016}; however, we have verified that fitting only to the SAURON wavelength region does not significantly affect our kinematics (\autoref{saurangecheck}). Lastly, the ATLAS\textsuperscript{3D}  kinematics were extracted using the MILES library of observed stars \citep{sanchezblazquez2006}, as opposed to the ELODIE library that we used.

\begin{figure*}
\begin{center}
	\includegraphics[trim = 0cm 9cm 0cm 0cm,scale=0.6]{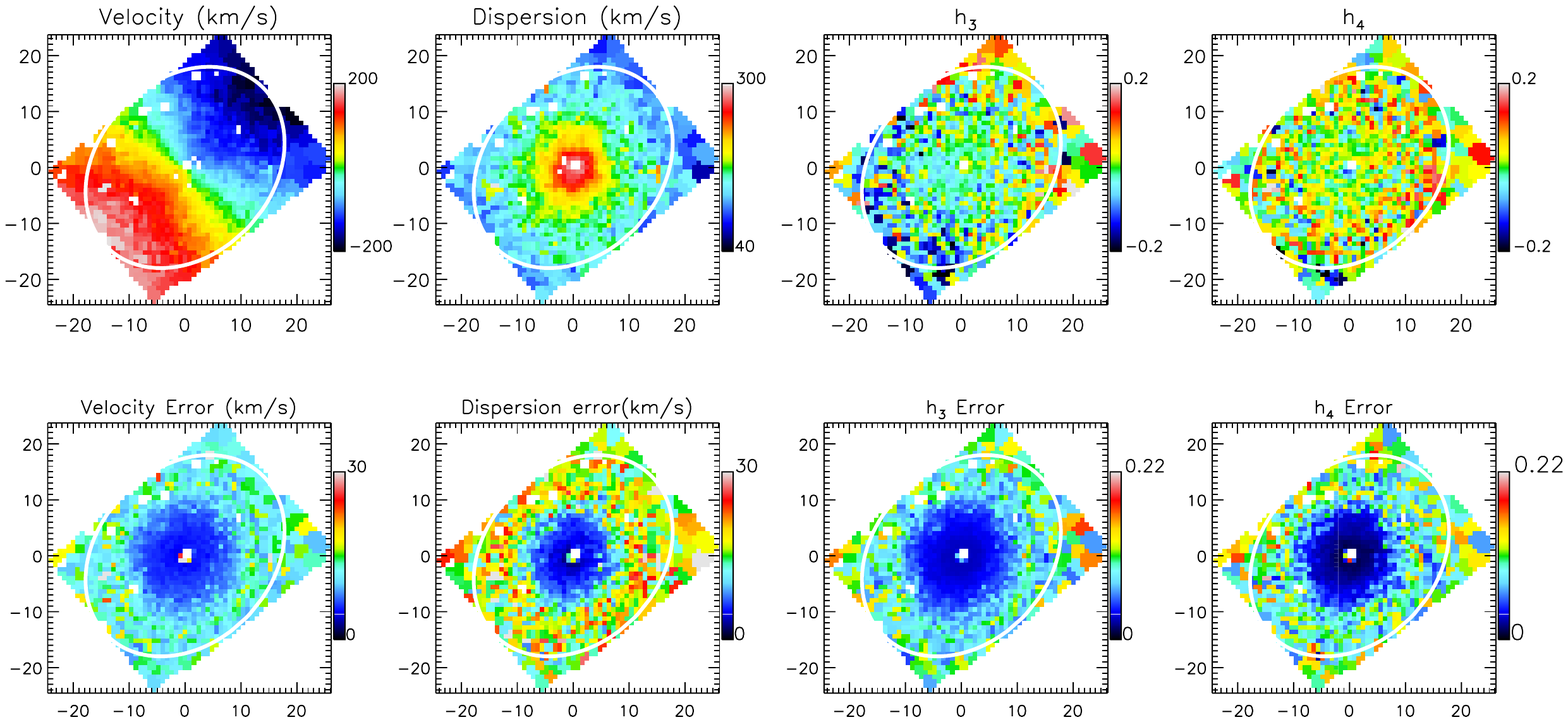}
	\caption{LOSVD and error maps that we re-measured from SAURON data, with the ELODIE library employed instead of the MILES library used in ATLAS\textsuperscript{3D}. The white ellipse shows the region enclosed by the innermost effective radius. Excluded datapoints are shown as white spaces.}
	\label{vdsauron}
	\end{center}
\end{figure*}

We therefore re-extracted the NGC 3998 SAURON kinematics from the public SAURON datacubes \footnote{Available from http://purl.org/atlas3d}. We used the ELODIE stellar templates as pPXF inputs, broadened to the SAURON spectral FWHM of 3.9\AA\ \citep{cappellari2011}. We again allowed for a 10th-degree additive polynomial correction and allowed the stellar template to vary between bins, for maximum consistency with the Mitchell kinematics, and we used the same bin locations as in the published ATLAS\textsuperscript{3D} data. We derived errors in pPXF by adding noise to spectra as described previously, and we again calculated additional systematic errors by re-fitting with MILES templates. We noted large offsets in a handful of central pixels, which possibly relates to template mismatch from the central LINER region; we added corresponding systematic error terms to the affected pixels individually. Over the remaining field of view, we derived systematic error terms of 4 km/s and 0.03 for the velocity  and $h_3$ terms respectively; we did not find it necessary to include systematic terms for the velocity dispersion or $h_4$, though a small number of bad pixels were identified in the same manner as with the Mitchell data. We present the resulting LOSVD and error maps in \autoref{vdsauron}. 

\begin{figure*}
\begin{center}
	\includegraphics[trim = 0cm 13cm 2cm 7cm, scale=0.8]{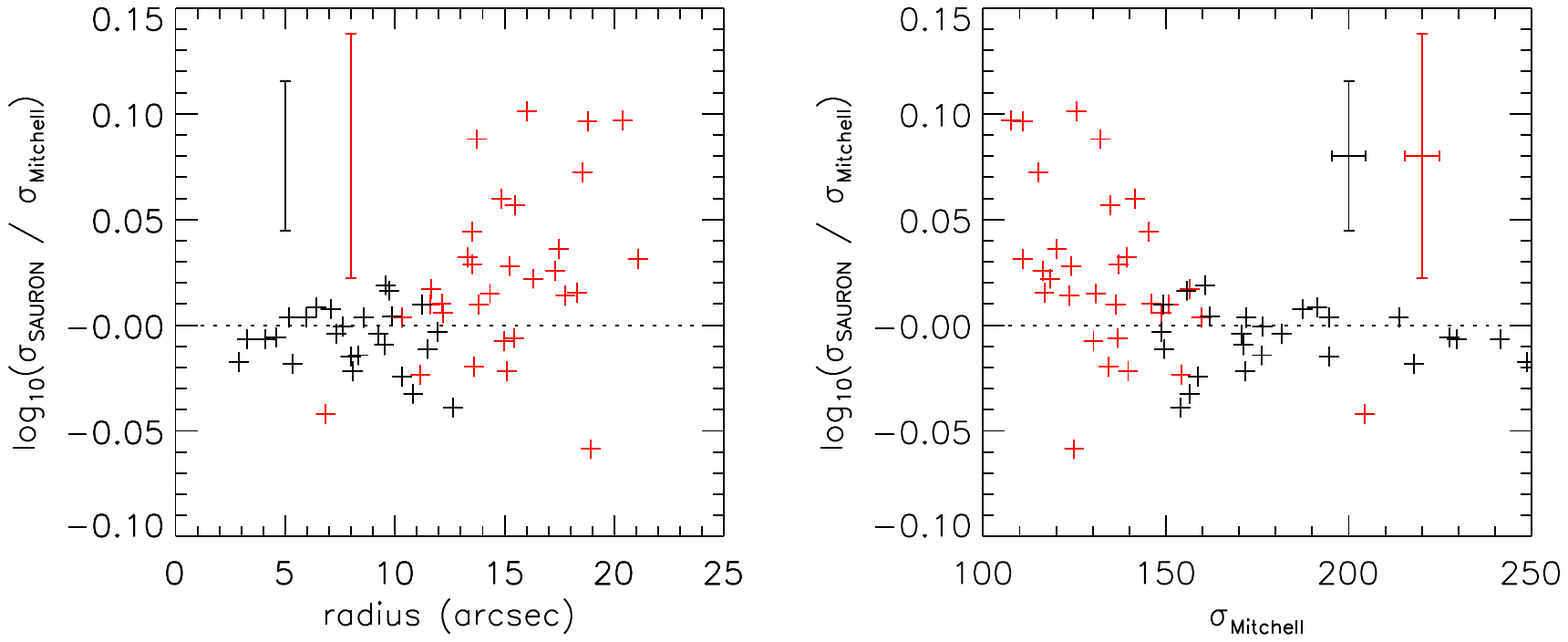}
	\caption{Ratio of the re-extracted SAURON velocity dispersions to those of Mitchell Spectrograph velocity dispersions, plotted against radius (left) and Mitchell dispersion (right). Lines and symbols are as in \autoref{losvdcompare1}. We find very little overall offset in unbinned SAURON regions, but once again see a number of large offsets where binning in the SAURON data becomes important.}
	\label{losvdcompare2}
	\end{center}
\end{figure*}

In \autoref{losvdcompare2}, we compare the dispersions from our Mitchell kinematics with those from the re-extracted SAURON results. Our own SAURON dispersions are generally slightly lower than reported in ATLAS\textsuperscript{3D}, but we continue to see some regions in the outskirts where the SAURON dispersions are higher than measured by the Mitchell. Overall, we find our kinematics to be slightly offset with respect to ATLAS\textsuperscript{3D}. 

We use our own SAURON kinematics in all subsequent analysis, in order for all of our data to have been extracted from the same stellar library for maximum consistency.

\section{Dynamical modelling}\label{mod}

We generated a series of orbit-based models of NGC 3998 using the Schwarzschild orbit superposition method \citep{schwarzschild1979}. We employed the method and code of \citet{vandenbosch2008},which allows for triaxial geometries and has previously been shown to accurately reproduce simulated ETGs \citep{vandeven2008,vandenbosch2009}. For a full description we refer the reader to \citet{vandenbosch2008}, but we provide a brief overview in section 4.1 along with details of our implementation. We describe the calculation of the spatial PSF in section 4.2.

\subsection{Method Overview}\label{method}

First, we construct a gravitational potential using a supplied galaxy surface brightness distribution, a supplied set of viewing angles and a supplied set of dark halo parameters. We construct an orbit library as follows. We first sample orbits along 21 equipotential shells logarithmically spaced between $0.5''$ and $294''$, and we sample over 7 radial and 7 angular directions for each shell. We then calculate an additional sets of orbits by dropping stars from successive equipotential curves, in order to ensure enough box orbits in our library over the galaxy's outer regions; these orbits are sampled along the same 21 equipotential shells and along $7 \times 7$ directions spaced evenly in terms of angles. We bundle together $3^3$ orbits with adjacent starting positions, producing a library of 18520 orbits overall.  We then determine the superposition of orbits that best fits the kinematics using a quadratic programming method to solve the least-squares problem. The size and shape of individual spectral bins is accounted for when fitting to the kinematics.

We parametrise the surface brightness using the Multi-Gaussian Expansion (MGE) method of \citet{emsellem1994}, which is implemented in the code of \citet{cappellari2002}. The MGE method models an observed surface brightness distribution as a set of two-dimensional Gaussian components, each of which may have unique peak values, widths and axial ratios. By deprojecting the galaxy for a set of viewing angles and assuming a constant mass-to-light (M/L) ratio, one can therefore obtain a density profile for the galaxy's visible component. For NGC 3998, we used the published I-band MGE model of \citet{walsh2012}, which was generated by combining \textit{HST} Wide-Field-Camera 2 (WFC2) imaging in the center with wide-field imaging from the Wide-Field InfraRed Camera (WIRCAM) on the Canadian-French Hawaiian telescope. The model is well-resolved in the centre and extends well beyond three effective radii. This MGE model has a photometric position angle (PA) of $-41.5^\circ$; this is slightly misaligned with our kinematic PA of $-44^\circ$, inferred from the method described in Appendix C of \citet{kraj2006}.

We parametrise the viewing angle using the three triaxial shape parameters $(q,p,u)$, which can the be related to the viewing angles $(\vartheta,\varphi,\psi)$ using equation (10) in \citet{vandenbosch2008}. In qualitative terms, $q$ represents the short-to-long axis ratio, $p$ the intermediate-to-long axis ratio, and $u$ the ratio between the projected and intrinsic MGE Gaussian widths along the galaxy major axis. For an axisymmetric galaxy, $p$ and $u$ are both equal to unity and $q$ is the only free shape parameter; in this case, $q$ is directly related to the galaxy viewing angle or inclination.  All three values correspond to the \textit{flattest} input Gaussian ($q' = 0.756$) of the input MGE surface brightness model; every MGE Gaussian component has a different projected axial ratio, and so will have different triaxial shape parameters at a given assumed viewing angle. 

We used kinematics from both the Mitchell Spectrograph and SAURON to fit the models.  We masked Mitchell fibres with centres less than 8$''$ from the galaxy centre and we masked SAURON pixels greater than 10$''$; this lets us combine the wide FOV of the Mitchell data with the higher spatial resolution of the central SAURON data, while limiting the overlap between datasets to avoid the fit being unfairly dominated by the galaxy centre. We brought the central values of the velocity and $h_3$ data to zero using the methods of \citet{kraj2006} and \citet{vandenbosch2010} respectively, which both calculate systemic components in the data as part of their routines.

We obtain the best-fitting superposition of orbits for each model using a quadratic programming solver \citep{gould2003}. We fit the orbits to our kinematics, with the intrinsic and projected model masses constrained to be within 2\% of the supplied MGE model at all times. We parametrise the quality of each fit using the $\chi^2$ statistic, which we define as

\begin{equation}\label{schwchi}
\chi^2= \sum_{i=1}^{N} \left(\frac {D_{mod,i} - D_i}{\Delta D_i}\right)^2
\end{equation}
\noindent
where $D$ signifies the values for each kinematic moment in turn, with $D_{mod}$ the values from a given model.

We set all confidence intervals using the expected standard deviation of the $\chi^2$ statistic, $\sqrt{2N_{obs}}$, where $N_{obs}$ is the total number of observables used to constrain the model parameters. This approach was first introduced in \citet{vandenbosch2008} and has since been validated in both \citet{vandeven2008} and \citet{vandenbosch2009}. It is preferable to using traditional sigma confidence intervals when $N_{obs}$ is large, as it better captures the uncertainties associated with the $\chi^2$ statistic itself. Here, $N_{obs} = 3866$ and so we define our confidence intervals using $\Delta\chi^2 = 87.9$. For comparison, the 3$\sigma$ confidence interval for four free parameters would be $\Delta\chi^2 = 16.25$.

We include a dark matter mass distribution by assuming a spherical NFW dark halo \citep{nfw}, in which the dark halo density profile is given as

\begin{equation}\label{dm1}
\frac {{\rho}(r)}{{\rho}_{crit}} = \frac {{\delta}_c}{(r/r_s)(1 + r/r_s)^2}
\end{equation}

where $r_s = r_{200}/c$ is a characteristic break radius and ${\rho}_{crit} = 3H^2/8{\pi}G$ is the critical density of the universe. c is the concentration parameter, while ${\delta}_c$ is the characteristic overdensity. This halo therefore has two free parameters: the concentration, and the halo mass $M_{200} = 200(4\pi /3)\rho r_{200}^3$. For the rest of this paper, we parametrise the halo mass in terms of $M_{200}/M_*$, where $M_*$ represents the stellar mass component of NGC 3998.

\subsection{PSF determination}\label{psf}

It is non-trivial to determine the PSF of our Mitchell observations, due to both the large individual optical fibres and the high number of individual datacubes that were combined during the data reduction process. We therefore elected to use the method of \citep{kraj2009}: a provided surface brightness model is convolved with a proposed PSF and then compared to an observed flux map, in order to optimise both the PSF parameters and the galaxy central position.

The models of \citet{vandenbosch2008} code parametrize the PSF as the sum of one or more Gaussian PSF components. We experimented with including two components, but found only one of these components to produce a non-negligible contributions; as such, we ultimately allowed for only a single Gaussian PSF component in the model. We allowed the position of the galaxy center to vary, and further allowed for a non-zero sky brightness to account for any residual sky contamination, producing four free parameters overall. We varied these parameters to determine the best-fitting spatial PSF and galaxy center, in each case convolving our MGE model with a given PSF Gaussian, and found a best-fit PSF of $\sigma = 1.40''$.

We also fitted a two-component PSF to the SAURON datacubes. The SAURON data cubes are reported to be positioned such that the galaxy nucleus is placed precisely on the IFU lenslets. We verified this by first letting the RA and DEC positions of the galaxy vary as before; we found both best-fitting values to be $ < 0.1''$ away from the stated numbers, and so we fitted a second time with the centre fixed to the ATLAS\textsuperscript{3D} position. Our best-fit PSF Gaussians have dispersions of $\sigma = 0.49''$ and $\sigma = 1.04''$ with relative weights of $0.7$ and $0.3$; since both Gaussians have non-negligible weights and imply the existence of extended PSF wings, we refrain from fitting the SAURON PSF with a single Gaussian component.

\section{Results}\label{res}

It is computationally expensive to fit for both the galaxy viewing angle and dark matter parameters simultaniously, due to the large number of free parameters involved. We therefore began by  estimating a reasonable set of NFW parameters for which to fit the stellar M/L and galaxy shape. We used the MGE model and best-fit M/L result of \citet{walsh2012} to derive a galaxy stellar mass of $7 \times 10^{10} M_\odot$, and we then estimated a dark-to-stellar mass fraction of 34.8 using the lowest-redshift fitting formula of \citet{leauthaud2012}. We then used the implied dark halo mass of $2.434 \times 10^{12} M_\odot$  to estimate an NFW concentration of $\log(c) = 0.867$ using the formula of \citet{sanchezconde2014}.

We then generated a grid of triaxial orbit models which sampled the shape parameters $(q,p,u)$ (for the \textit{flattest} MGE Gaussian component) and the stellar M/L. We sampled $q$ down to a low minimum value of 0.06, motivated by the FR shape results of \citet{weijmans2014}, and we sampled $p$ down to a value of 0.92. We used all possible values of $u$, as limited by the values of $q$ and $p$ \citep[see, e.g.,][for an explanation]{vandenbosch2008}. We sampled 16 I-band stellar M/L values evenly between 3.1 and 6.1. We found a galaxy shape of $(q,p,u) = (0.06_{-0.00}^{+0.18},0.98_{0.004}^{0.003},0.99_{-0.001}^{+0.002})$. The lower bound on $q$ corresponds to the lowest tested value and so is not a true lower limit. The narrow bounds on $p$ and $u$ are due the small misallignment between the photometric and kinematic PAs discussed previously, which produces a kinematic misallignment between models and data (and so a large $\chi^2$ increase) unless $p$ and $u$ take very specific values. 

The shape parameters $(q,p,u)$ correspond to the \textit{flattest} component of the input MGE model. We calculate the resulting shape of the overall model by first deprojecting it according to $(q,p,u)$ and then fitting spheroids of constant luminosity density. We thus find a minor-to-major axis ratio at $1R_e$ of $q_{R_e} = 0.44_{-0.00}^{+0.05}$. We conclude from this that NGC 3998 is a flattened and near-oblate galaxy.

We next generated a series of models in which we fitted for the NFW dark halo mass, the stellar M/L and the axis ratio $q$. We fixed $p$ and $u$ to values of 0.98 and 0.99 respectively, as per the discussion on the previous paragraph. We used $q$ values between 0.06 and 0.38, in steps of 0.08, to ensure that all reasonable $q$ values were covered given the results of the previous model grid. We fixed the halo concentration $c$ to the same value as before; this is because the NFW scale radius will be significantly beyond our FOV for most reasonable halo parameters, making the dark mass and concentration highly degenerate \citep[see, e.g.][for a similar argument]{yildirim2016}. We used a range of $\log_{10}(M_{200}/M_{*})$ from 0.01 to 4.09, with models without dark matter also included. We sampled the same M/L values as before. 

\begin{figure}
\begin{center}
	\includegraphics[trim = 5cm 3cm 5cm 3cm, scale=0.5]{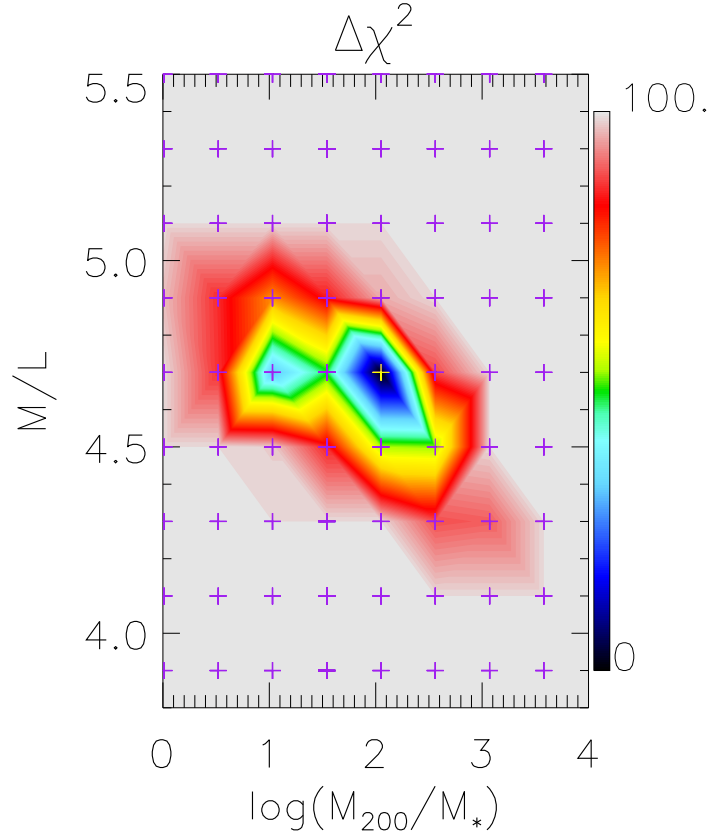}
	\caption{Map of $\Delta(\chi^2)$ in terms of the dark matter mass fraction and the stellar M/L, with the axis ratio q marginalised over. The yellow point represents the best-fitting model.}
	\label{chi2grid}
	\end{center}
\end{figure}

\begin{figure}
\begin{center}
	\includegraphics[trim = 5cm 3cm 5cm 3cm, scale=0.5]{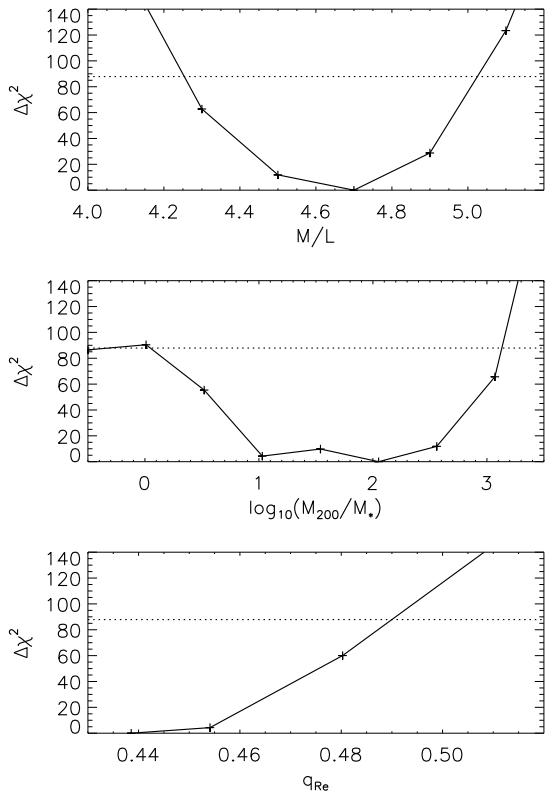}
	\caption{$\Delta\chi^2$ plotted as a function of the I-band M/L (top), the NFW dark halo mass (middle) and the axial ratio $q_{R_e}$, with all other parameters marginalised over in each case. The leftmost point on the middle window is for the DM-free case. The horizontal dashed line signifies $\Delta\chi^2 = 87.9$, which marks the bounds of our confidence region and is set by the standard deviation of the $\chi^2$ statistic. We obtain relatively tight constraints on the stellar M/L and we obtain tight upper limits on $q_{R_e}$ and the dark halo mass.}
	\label{chi2prof}
	\end{center}
\end{figure}

We present the $\Delta\chi^2$ for the above-described model in terms of $M_{200}$, M/L in \autoref{chi2grid}, in which the galaxy shape has been marginalised over, and we present separate $\Delta\chi^2$ profiles for  $M_{200}$, M/L and $q_{lum}$ in \autoref{chi2prof}. Our best-fit model has a $\chi^2/DOF$ of 1.11. We find $M/L = 4.7_{-0.45}^{+0.32}$, with a dark halo mass $\log_{10}(M_{200}/M_{*}) \leq 3.13$. We find no firm lower limit on the dark halo mass, as the best-fitting DM-free case falls within our $\chi^2$ confidence limit.

Marginalising over M/L and the dark matter mass, we find $q_{R_e} = 0.44_{-0.00}^{+0.05}$, where the minimum is again set by the limits of our sampling; this corresponds to an upper limit of $42.5^\circ$ on the viewing angle $\theta$. If we instead consider models with no dark matter only, we find $q_{R_e} = 0.48_{-0.04}^{+0.04}$ ($\theta < 43.4^\circ$). Thus, the $q_{R_e}$ confidence region shows a slight dependence on the presence of dark matter, with the best fit $q_{R_e}$ value higher in the DM-free case.

\begin{figure}
\begin{center}
	\includegraphics[trim = 5cm 13cm 5cm 4cm, scale=0.5]{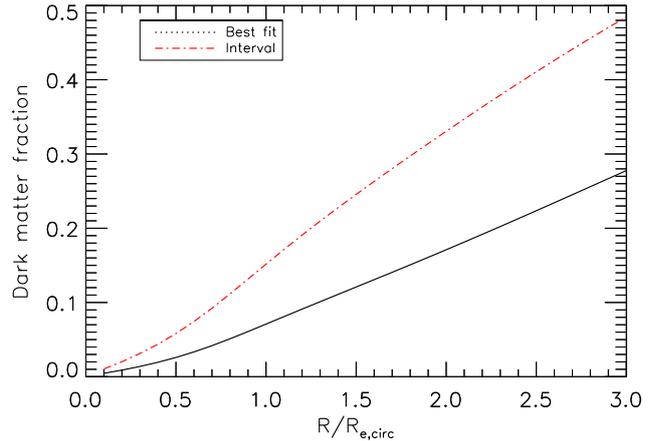}
	\caption{Profile of best-fit enclosed dark mass fraction out to 3$R_{e,circ}$(black line), along with the upper limit (dash-dotted red line). We show no lower limit because our best-fit DM-free model falls within our $\chi^2$ confidence limit. $R_{e,circ}$ is the circularised effective radius from table 1 of \citet{cappellari2013}. We find tight upper limits on the dark matter content, with the models preferring little dark matter within 1$R_{e,circ}$.} 
	\label{dmfrac}
	\end{center}
\end{figure}

In \autoref{dmfrac} we plot the inferred best-fit dark matter fraction along with the associated confidence interval. We define the dark matter fraction as the fraction of dark mass within a sphere of given radius, using a circularised MGE surface brightness model. We find a best-fitting dark matter fraction of $(7.1^{+8.1}_{-7.1})\%$ within $1R_{e,circ}$, where $R_{e,circ}$ is the circularised effective radius of $23.99''$ reported in table 1 of \citet{cappellari2013}, which is in agreement with the dark fraction of $(15 \pm 6)\%$ reported in \citet{cappellari2013a}. Thus, we find that low dark mass fractions are preferred by our models.

\section{Discussion}\label{disc}

In the previous section, we derived best-fit parameter ranges for NGC 3998's shape, I-band stellar M/L, and NFW dark halo parameters. We found NGC 3998 to be a near-oblate, face-on fast rotator, in agreement with the flattening inferred for this galaxy in \citet{cappellari2013}. We obtain tight constraints on the intrinsic galaxy shape, unlike in \citet{walsh2012}.  Our best-fit I-band M/L is also somewhat lower than reported in \citet{walsh2012}, though it is consistent within the errors. The latter point can immediately be understood as the consequence of including a dark halo, while the use of a wide two-dimensional field-of-view enables significantly tighter constraints on the shape parametes than was previously possible. We find an axial ratio $q_{R_e} = 0.44_{-0.00}^{+0.05}$ from our models, which corresponds to a viewing angle $\theta$ lesser than  $42.5^\circ$; this is consistent with the inclination of $(38\pm 5)^\circ$ reported in \citet{cappellari2013}.

Our 1$R_{e,circ}$ dark matter fraction of $(7.1^{+8.1}_{-7.1})\%$ is lower than commonly reported for ETGs from simulations or observations \citep[e.g.][]{weijmans2009,barnabe2011,remus2013,wu2014}, but is consistent with the fraction of $(15 \pm 6)\% $ reported in \citet{cappellari2013a} and is also well within the scatter of low dark fractions reported by \citet{cappellari2013}.    We note that the inferred stellar mass for NGC 3998 is close to the $M_*-M_{halo}$ pivot mass of $4.5 \times 10^{10} M_\odot$ inferred from weak lensing studies of nearby galaxies \citep{leauthaud2012}, meaning that a lower-than-average dark mass fraction is perhaps not surprising. 

\citet{cappellari2013a} report an r-band M/L of $6.58 \pm 0.39$ for this galaxy, from stellar population modelling assuming a Salpeter IMF slope. We convert this to an I-band M/L of $4.72 \pm 0.28$ using 

\begin{equation}
(M/L)_I = (M/L)_r \times 10^{\frac{(I-r)_{gal} - (I-r)_\odot}{2.5}}
\end{equation}

\noindent
where we obtain a galaxy color $(I-r)_{gal} $ of $-0.48$ by comparing the \citet{walsh2012} MGE model with the r-band MGE model from ATLAS\textsuperscript{3D} \citep{scott2013} out to $50''$, and where we assume a Sun color of $(I-r)_\odot = - 0.12$ \citep[and references therein][]{blanton2007}. Our I-band M/L of $4.7_{-0.45}^{+0.32}$ inferred from Schwarzschild modelling is therefore in excellent agreement with the \citet{cappellari2013a} stellar population models. For models without dark matter, we find an I-band M/L of $4.9_{-0.25}^{+0.27}$, which is also within the uncertainty of the \citet{cappellari2013a} value. Thus, we find that the \citet{cappellari2013a} models are consistent with a low dark fraction for this galaxy.

\begin{table}
\begin{center}
\begin{tabular}{|c|c|c|c|c|}
\hline 
Model & $q$ & M/L & $\log_{10}\left(\frac{M_{200}}{M_{*}}\right)$ & $f_{DM}(R_{e,circ})$\\ 
\hline 
Best fit & 0.06 &  4.7 & 2.05 & 7.1\%\\ 
\hline 
DM-heavy & 0.06 &  4.1 &  3.58 & 20.2\%\\ 
\hline
DM-free & 0.22 & 4.9 & N/A & 0\%\\ 
\hline
\end{tabular} 
\end{center}
\caption{Summary of the three models shown in Figures 13-16 and discussed in the text. $f_{DM}(R_{e,circ})$ denotes the dark fraction within one (circularised) effective radius.}
\label{tab2}
\end{table}

\begin{figure*}
\begin{center}
	\includegraphics[trim = 2cm 0cm 0cm 9cm,scale=0.85]{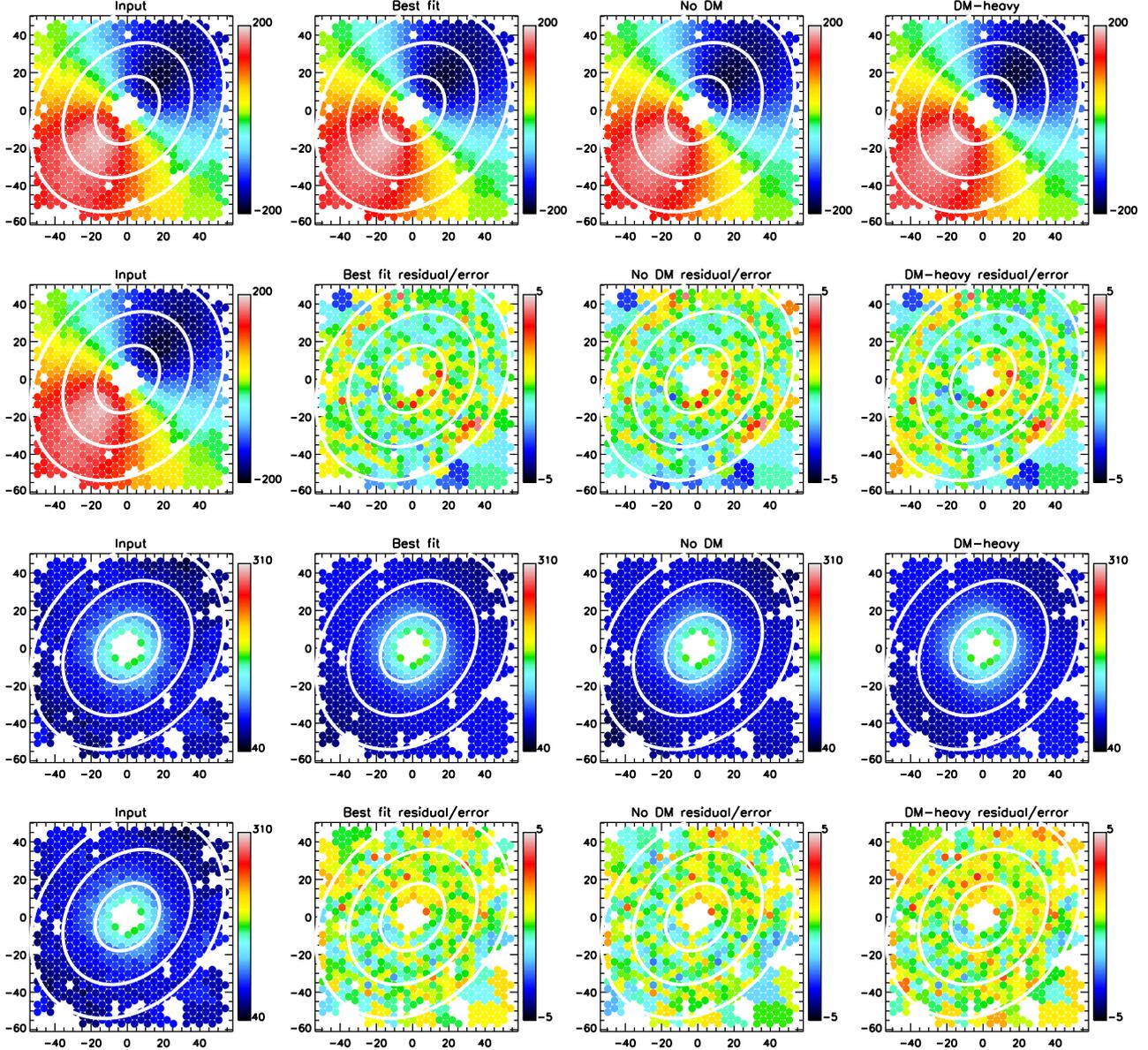}
	\caption{From top to bottom: input Mitchell velocity and model velocities, (model - data) velocity residual/error, input and model dispersions and (model - data) dispersion. We summarise the properties of the three selected models in \autoref{tab2}. Only unmasked bins are shown.} 
	\label{modelsvsig}
	\end{center}
\end{figure*}

\begin{figure*}
\begin{center}
	\includegraphics[trim = 2cm 0cm 0cm 9cm,scale=0.85]{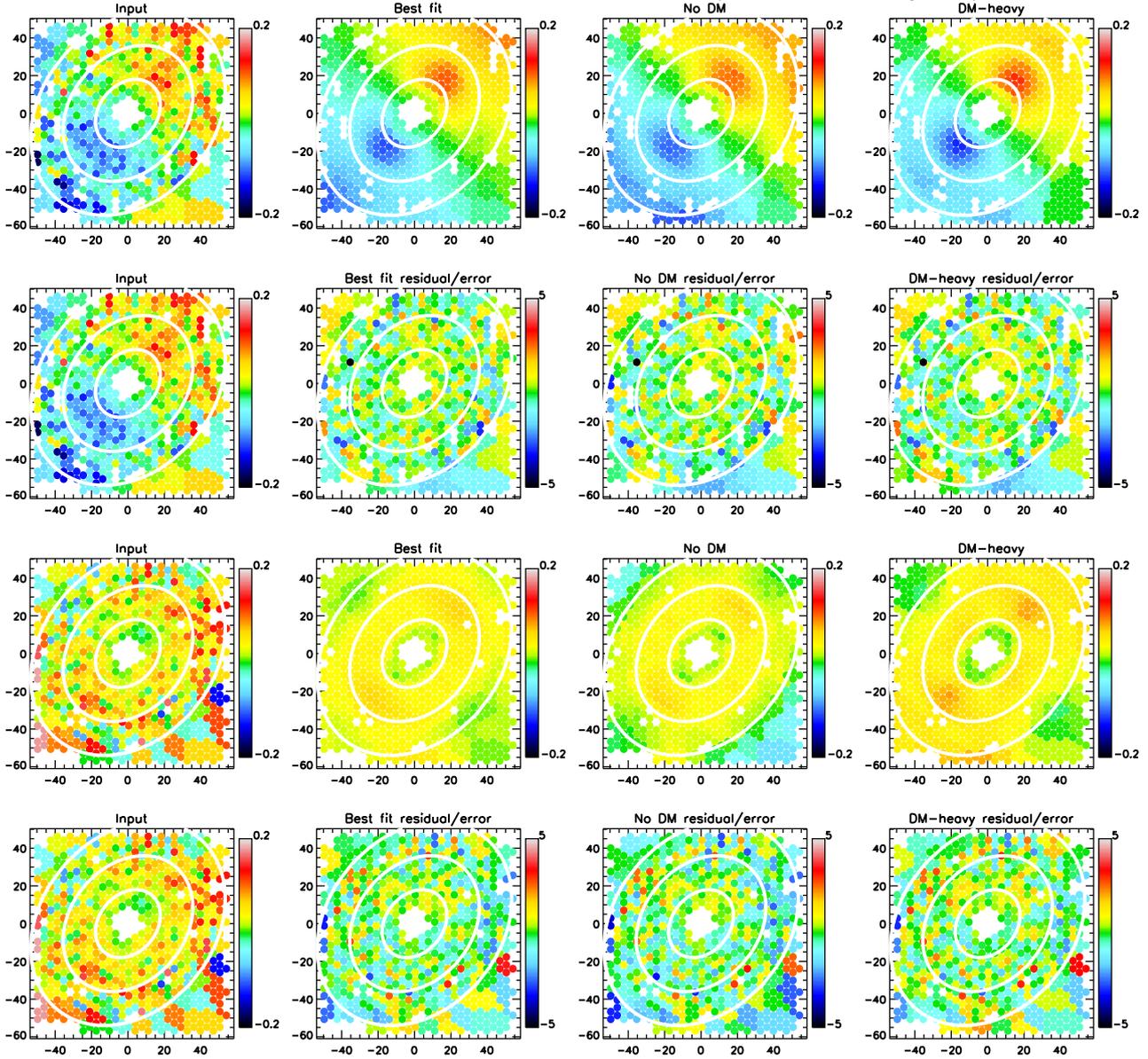}
	\caption{As in \autoref{modelsvsig}, but for $h_3$ and $h_4$.} 
	\label{modelsh3h4}
	\end{center}
\end{figure*}

\begin{figure*}
\begin{center}
	\includegraphics[trim = 2cm 0cm 0cm 9cm,scale=0.9]{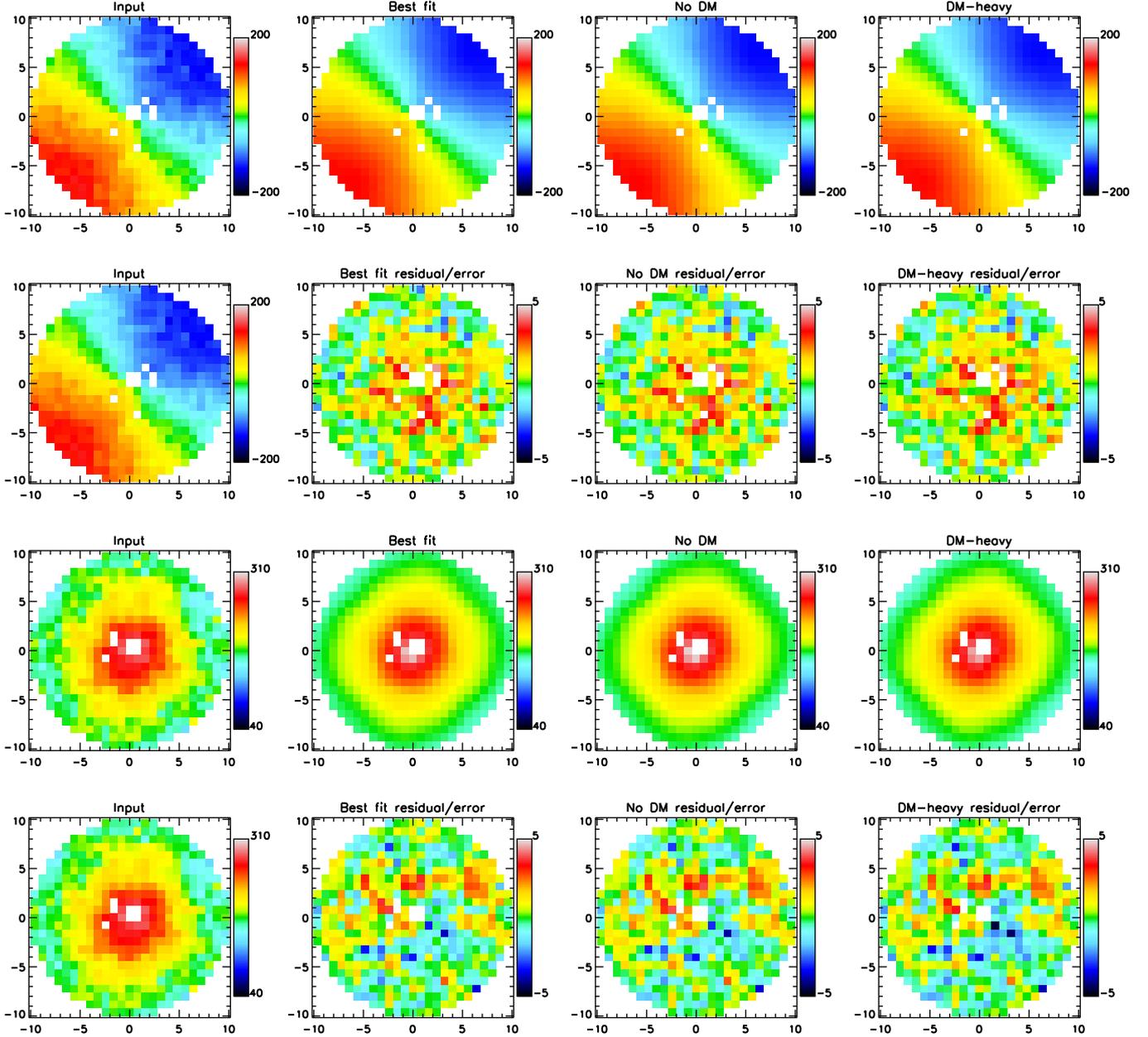}
	\caption{As in \autoref{modelsvsig}, but for the SAURON data.} 
	\label{saumodelsvsig}
	\end{center}
\end{figure*}

\begin{figure*}
\begin{center}
	\includegraphics[trim = 2cm 0cm 0cm 9cm,scale=0.9]{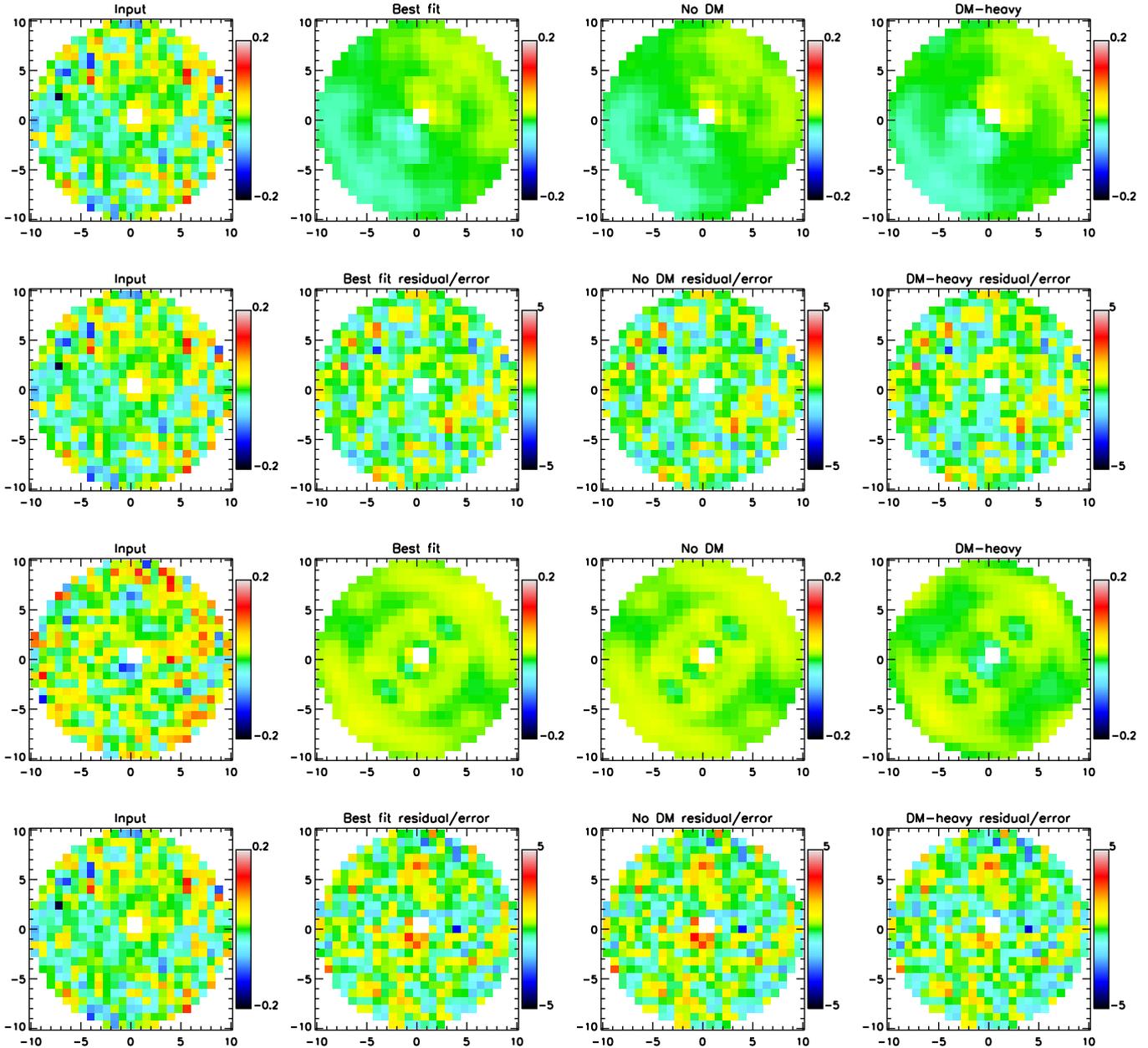}
	\caption{As in \autoref{modelsvsig}, but for $h_3$ and $h_4$ from the SAURON data.} 
	\label{saumodelsh3h4}
	\end{center}
\end{figure*}

\begin{figure}
\begin{center}
	\includegraphics[trim = 1.5cm 2cm 0cm 18cm,scale=0.85]{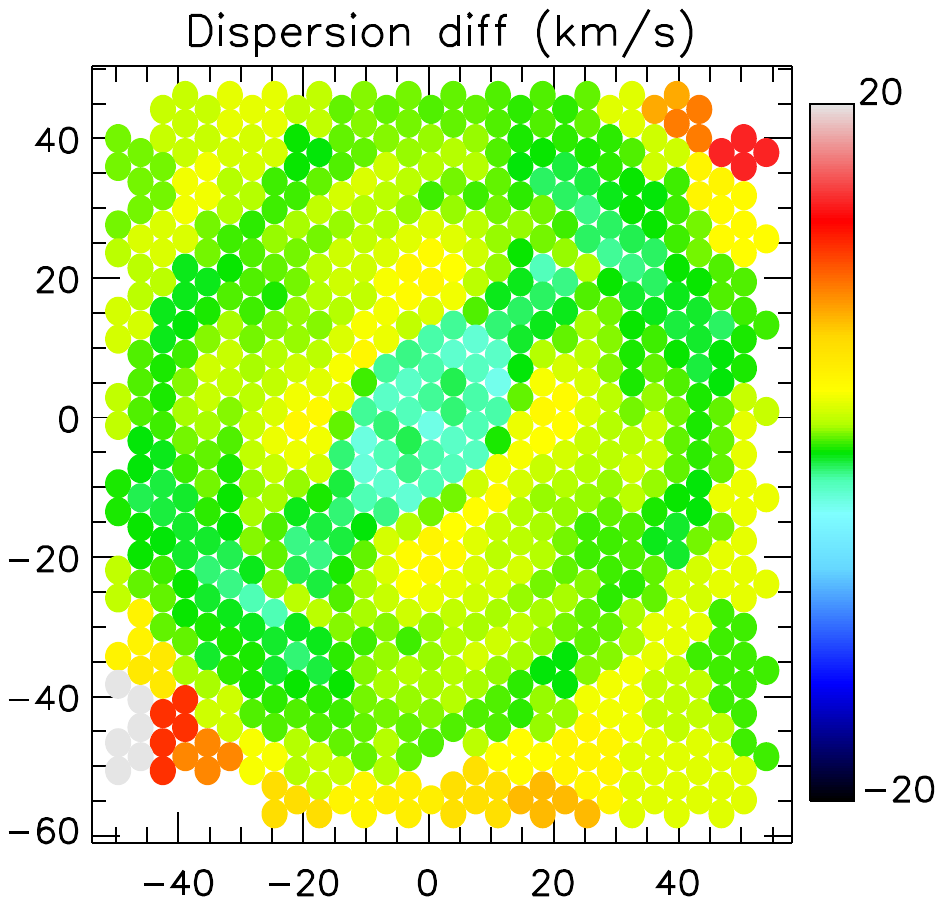}
	\caption{Difference in line of sight velocity dispersion between the selected DM-heavy model and the best-fit model, where a positive number indicates the former model being higher. The dispersion is somewhat higher over much of the FOV beyond $1 R_e$, which is the main cause of the DM-heavy model's higher $\chi^2$.} 
	\label{modeldif}
	\end{center}
\end{figure}

In \autoref{modelsvsig} and \autoref{modelsh3h4}, we compare our input Mitchell kinematics maps to three of our orbit models: our best-fit model, a dark matter heavy model and our best-fitting dark-matter-free model. We show the same comparisons for the SAURON dara in \autoref{saumodelsvsig} and \autoref{saumodelsh3h4}. We summarise the properties of the selected models in \autoref{tab2}. The DM-free model is within our $\chi^2$ confidence criterion, while the DM-heavy model is significantly beyond it ($\chi^2 - \chi^2_{lim} = 164.1$). The DM-free model looks very similar to the overall best-fit model, which is unsurprising given the low dark matter content that we infer. The DM-heavy model also looks similar over most of the FOV, though its velocity dispersions are generally somewhat higher in the outskirts than for the best-fit model (\autoref{modeldif}); this is the main cause of the $\chi^2$ difference between the two models.

We have seen above that higher dispersion values in the outskirts correspond to larger dark matter fractions. Were our dispersions to be underestimated, then this could bias our results towards lower dark matter fractions. We discussed the robustness of our kinematics in \autoref{kin}, during which we noted that our Mitchell velocity dispersions are somewhat lower away from the centre than would be implied from ATLAS\textsuperscript{3D}. However, we found good agreement in the two instruments' dispersions within the \textit{unbinned} ATLAS3D region, and we have also verified that fitting Mitchell data with MILES does not significantly affect the inferred offset between the instruments. We also note that pPXF fits with MILES produce somewhat higher dispersions in the outskirts of the Mitchell data than fits with ELODIE, with a median $7.8\% $ difference for ELODIE dispersions below 70 km/s;  however, this is approaching the intrinsic resolution of the MILES library (\~ 60 km/s) and so we view the ELODIE kinematics as more reliable over this region. Since our dark matter fraction is consistent with the ATLAS\textsuperscript{3D} value, we do not explore this point further.  

\begin{figure}
\begin{center}
	\includegraphics[trim = 5cm 13cm 3cm 3.5cm,scale=0.55]{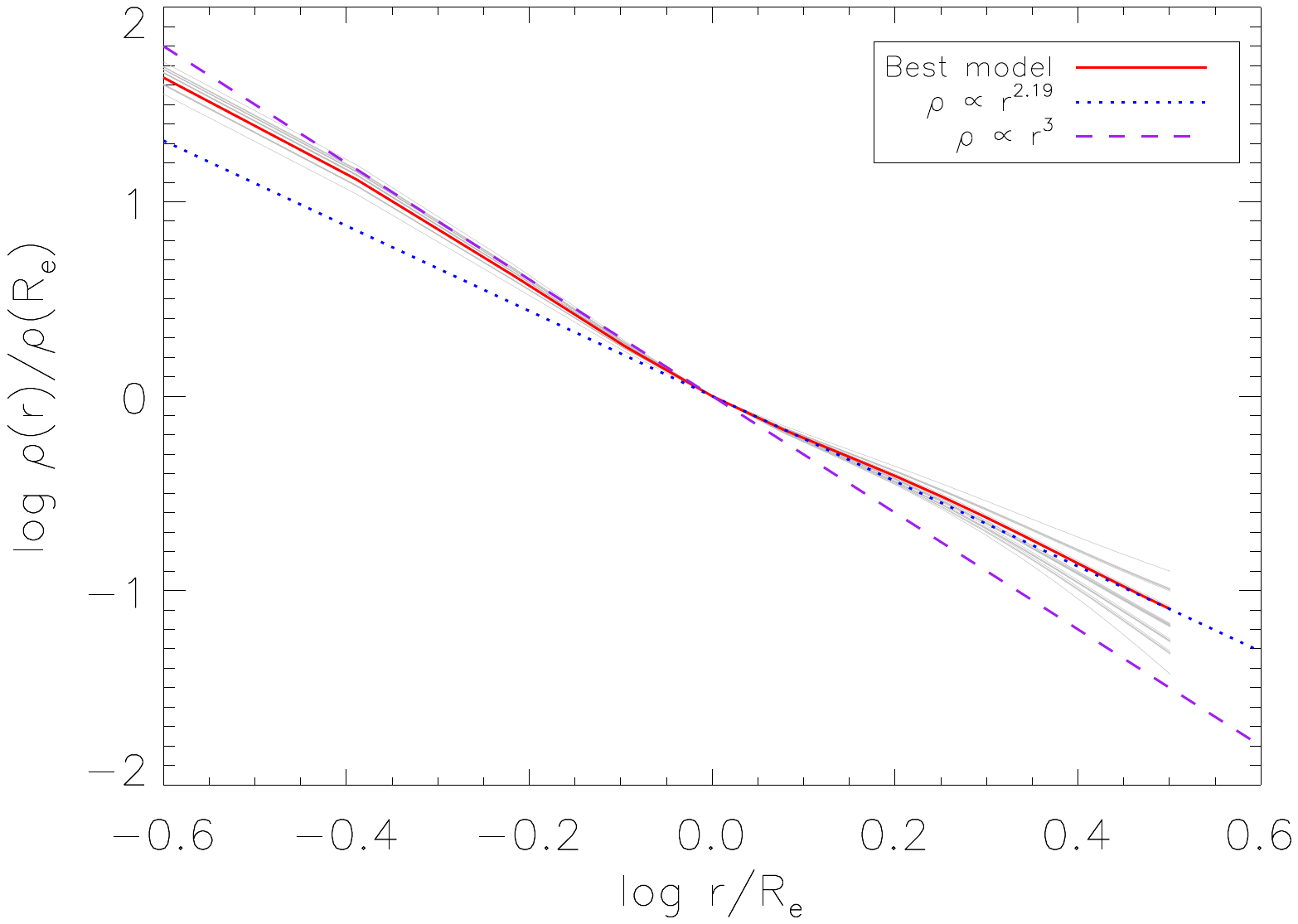}
	\caption{Total (dark plus baryonic) density profiles of all allowed Schwarzschild models, with lines for $\rho \propto r^{2.19}$ \citep{cappellari2015} and $\rho \propto r^{3}$ shown for comparison. We find the profiles to follow near-isothermal behaviour beyond the central effective radius, in good agreement with \citep{cappellari2015}.}
	\label{totalrho}
	\end{center}
\end{figure}

In \autoref{totalrho}, we plot the total (dark plus baryonic) mass density profiles of all allowed Schwarzchild models. We find the models to follow near-isothermal profiles beyond the central effective radius, in good agreement with the average logarithmic slope of $2.19 \pm 0.03$ reported from the dynamical models of \citet{cappellari2015}. The density slopes within the central effective radius are somewhat steeper, and resemble the stars-only density profiles of that same paper; this is unsurprising, given the low central dark matter fractions of all our our allowed models.

\begin{figure}
\begin{center}
	\includegraphics[trim = 1cm 1cm 0cm 9.5cm,scale=0.45]{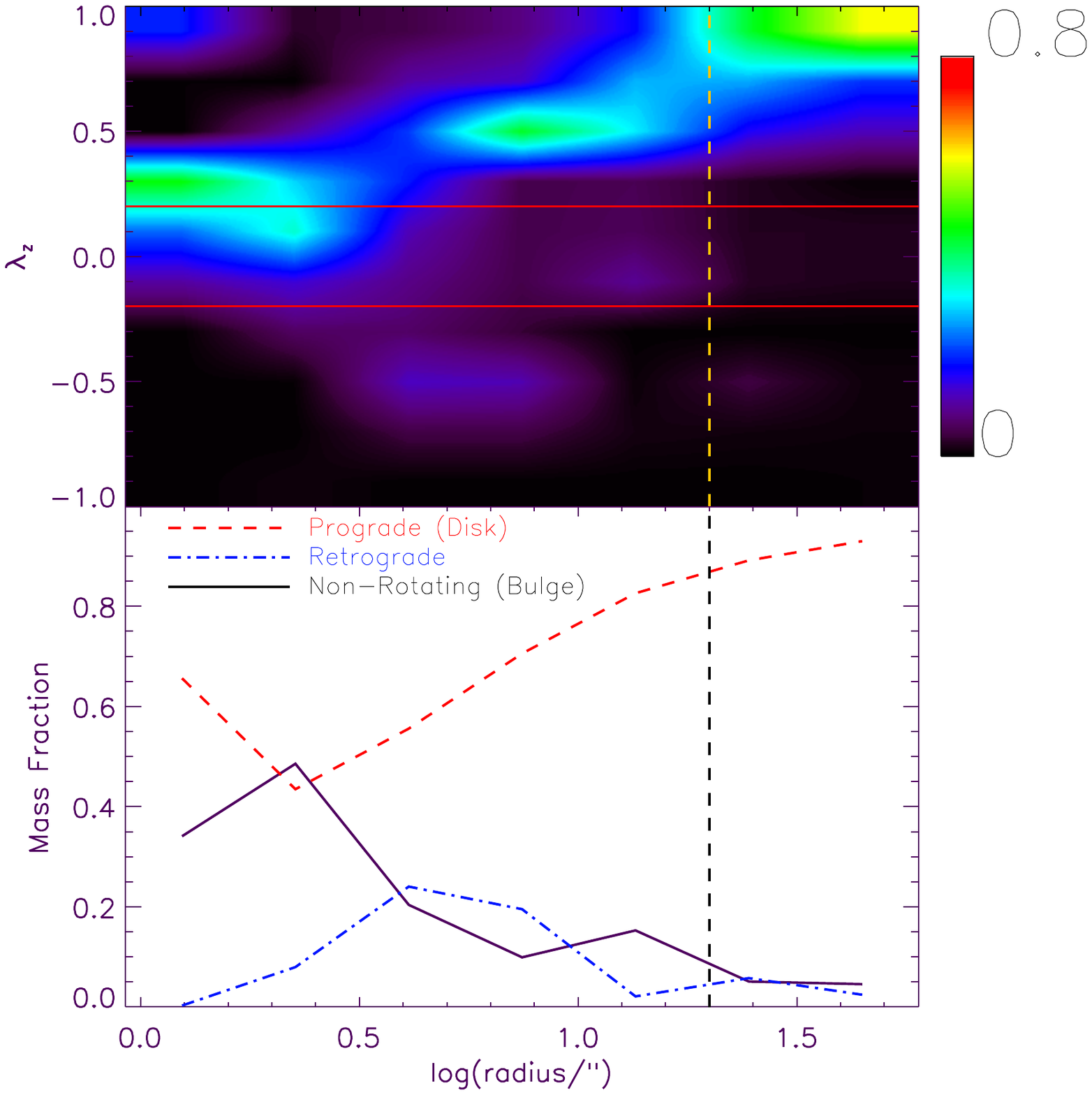}
	\caption{Mass distribution of orbits for the best-fitting orbit model, plotted as a function of average radius and spin. The distribution on the top panel is normalised per unit radius The vertical dashed lines mark 1 $R_e$. A bulge-disc separation is evident, with the disc coming to dominate in the outer observed regions.}
	\label{momentummap}
	\end{center}
\end{figure}

\begin{figure}
\begin{center}
	\includegraphics[trim = 1cm 1cm 0cm 9.5cm,scale=0.45]{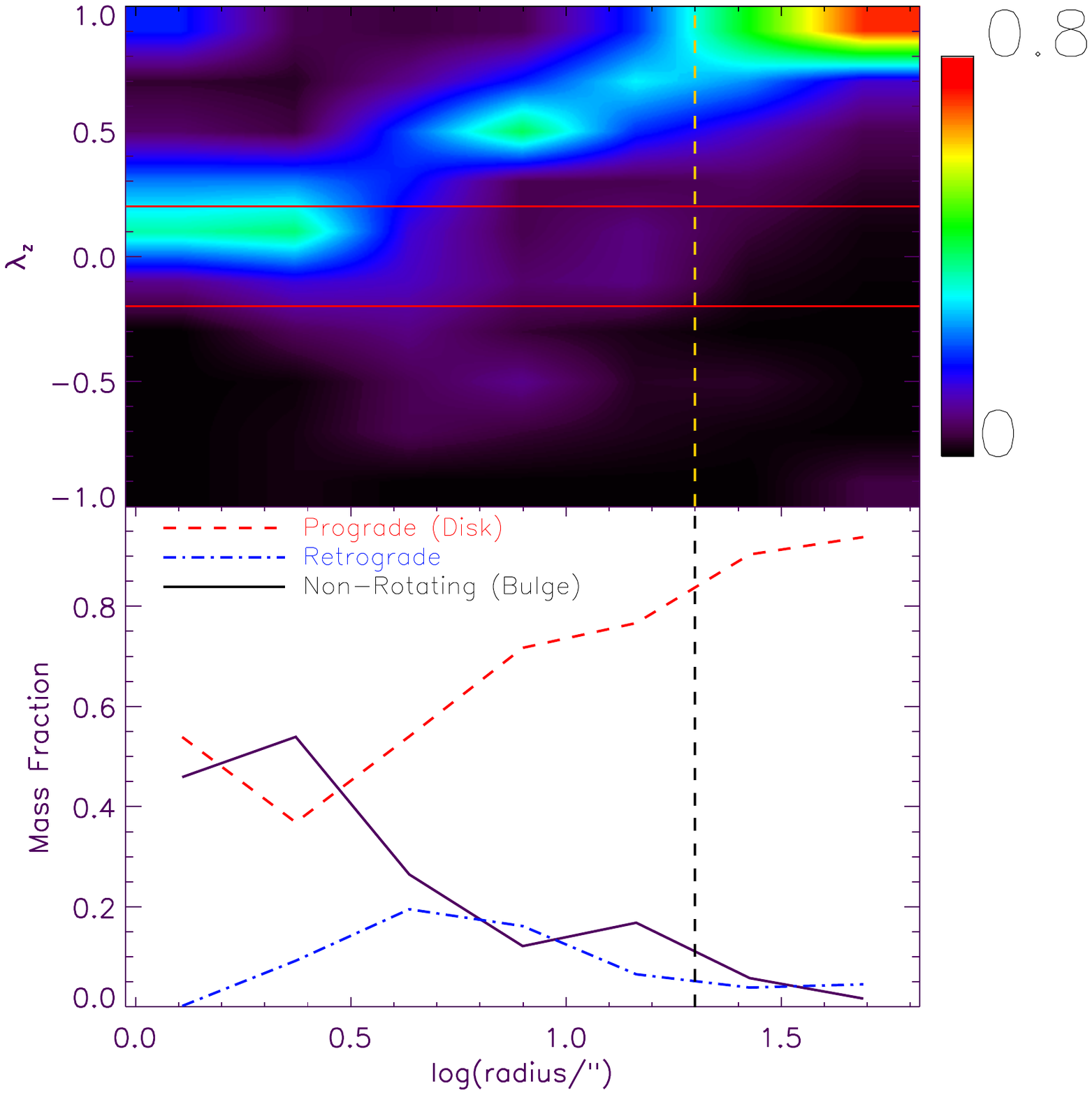}
	\caption{Same as \autoref{momentummap}, but for the DM-free model.}
	\label{momentummapnodm}
	\end{center}
\end{figure}

In \autoref{momentummap} we present the distribution of orbits in our best-fitting orbit model model as a function of spin parameter $\lambda_z$ and position.  We find evidence of both a slow-rotating bulge component and a fast-rotating disc, with the disc becoming increasingly dominant beyond the central half-light radius. We also find a non-negligible counter-rotating component within the central 10 arcseconds, which is not visible in the SAURON or Mitchell kinematics maps. Within $1R_e$, we find the orbital distribution to consist of 61.3\% prograde orbits, 26.1\% non-rotating ($\lambda_z < 0.2$) orbits and 12.6\% retrograde orbits; within $3 R_e$, we find 70.5$\%$ prograde orbits, 19.5\% non-rotating orbits and 10.0\% retrograde orbits. We show the distribution of orbits for the selected DM-free model in \autoref{momentummapnodm}; we find that the distribution is qualitatively similar to that seen in \autoref{momentummap}, with a bulge and disc component present along with a non-negligible counter-rotating component. We find counter-rotating components of similar mass fraction to be present in all models allowed by our $\chi^2$ criterion; as such, this appears to be a necessary component to successful model fits. Overall, these figures provide an excellent illustration of how multiple datasets may be employed to study an object in detail. 

\begin{figure}
\begin{center}
	\includegraphics[trim = 2cm 1cm 0cm 19cm, scale=0.8]{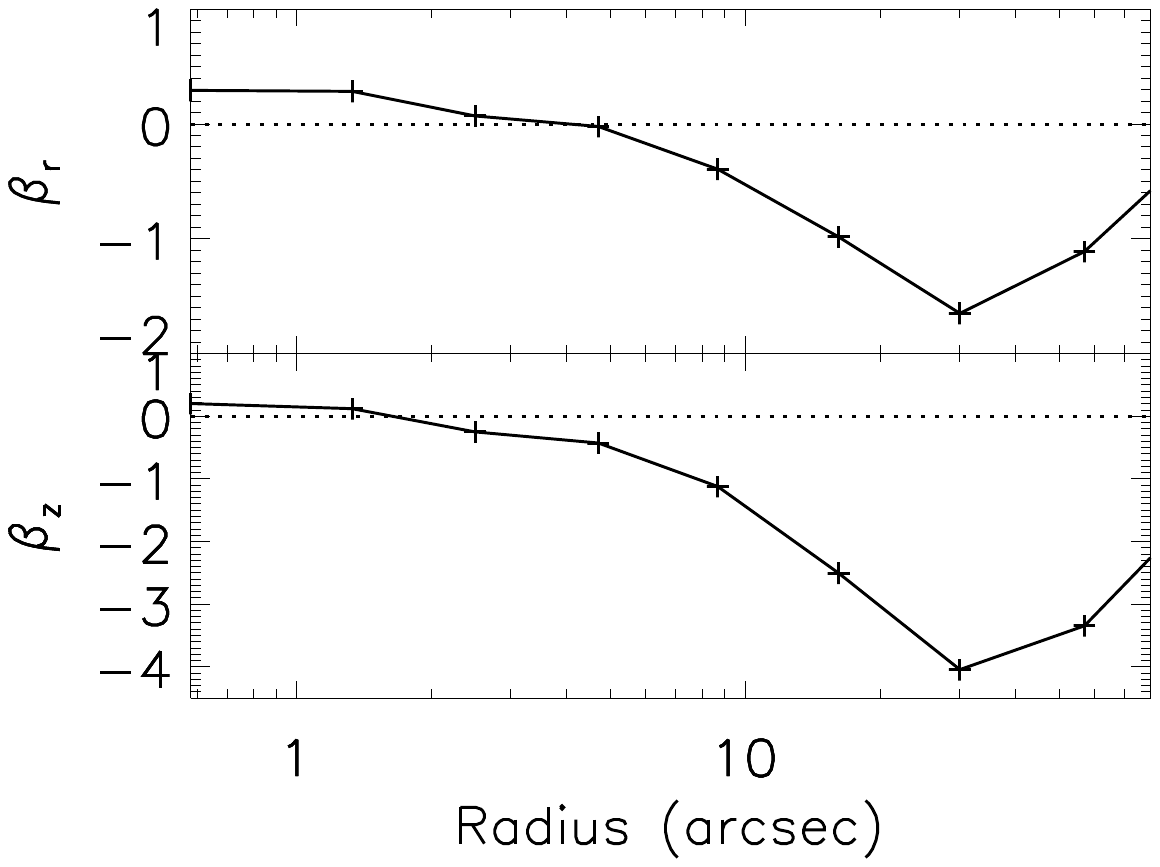}
	\caption{Orbital anisotropy parameters $\beta_r$ (top) and $\beta_z$ (bottom) for the best-fitting model, plotted as a function of radius along the galaxy major axis. The dotted lines represent isotropy. We find the orbital anistropy to vary signficantly as a function of position.}
	\label{betaz}
	\end{center}
\end{figure}

In \autoref{betaz} we plot the orbital anisotropy parameters $\beta_r = 1 - (\sigma_t/\sigma_r)^2$ and $\beta_z = 1 - (\sigma_z/\sigma_R)^2$ as a function of radius, as inferred from our best-fit model.  We define the tangential velocity dispersion $\sigma_t$ as $\sigma_t^2 = (\sigma_\theta^2 + \sigma_\phi^2)/2$, where ($r$,$\theta$,$\phi$) are the standard spherical coordinates. $(R,z)$ are cylindrical coordinates, with $\sigma_z$ therefore denoting the dispersion out of the plane of the disc. We find our best-fit model to be radially anisotropic ($\beta_r > 0$) in the centre, similar to the results of \citet{walsh2012}, and we find tangential anisotropy further out; this can be understood as a result of the bulge-disc separation discussed above. We find $\beta_z$ to vary strongly as a function of position; this is different from the assumption of constant $\beta_z$ employed in the axisymmetric jeans models of \citet{cappellari2013a}, even within the region covered by SAURON data. 

An important caveat here is that the low inclination of the galaxy introduces a significant degeneracy in the MGE deprojection, as small changes to the (projected) MGE surface brightness model could potentially produce large changes in the deprojected mass distribution. Such changes would affect the inferred extent and flattening of individual galaxy components. Since rounder galaxy models are known to produce more radial anisotropies \citep[e.g.][]{magorrian2001, cappellari2008}, changes to individual components can be expected to affect the anisotropy similarly; this, then, is an alternative explanation for our finding of strong tangential anisotropy. A similar analysis on the outskirts of more edge-on galaxies would be revealing in this regard. 

From the above discussion, our modelling implies NGC 3998 to be disc-dominated over much of the Mitchell FOV with only a small proportion of radial orbits in the outskirts. Late-time accretion of stars is expected from simulations to produce an excess of radial orbits beyond the central effective radius \citep[e.g.][]{oser2010,wu2014}, while tangential anisotropy is only observed in simulated galaxies with large fractions of in-situ formed stars \citep{wu2014}; we therefore infer that late-time accretion did \textit{not} play a major role in this galaxy, in agreement with the proposed FR formation path of \citet{cappellari2013b} (in which only SRs have experienced late dry accretion). Simulations further suggest tangential anisotropy to indicate galaxies which have experienced a recent ($z < 2$) dissipational event \citep{rottgers2014}, with the tangential anisotropy a result of the large fraction of in-sity formed stars. Our inferred galaxy shape of $q_{R_e} = 0.44_{-0.00}^{+0.05}$ is also consistent with a past dissipational event, from the cosmological simulations of \citet{naab2014}. 

Overall, our models support the FR formation picture presented in \citet{cappellari2013b}. Our models imply little late-time accretion for this galaxy, which disfavours the two-phase evolutionary history associated with SRs. Our models are, however, consistent with some past dissipational event having occurred. Given the apparent lack of late-time accretion, we would not expect to find multiple components to this galaxy's stellar population; rather, we predict this galaxy to be dominated by an old population of internal origin. Such a prediction is also supported by the simple stellar population modelling of \citet{mcdermid2015}, who report an old ($11.27 \pm 1.95$) stellar population of near-solar metalicity ($[Z/H] = -0.04 \pm 0.05$) from simple stellar population fits to the central effective radius.

\section{Summary and Conclusion}\label{sum}

We have presented an orbit-modelling analysis of the lenticular galaxy NGC 3998, with kinematics obtained from both Mitchell Spectrograph observations as well as from archival SAURON observations taken as part of ATLAS\textsuperscript{3D}. Our modelling implies NGC 3998 to be a near-oblate and flattened galaxy, consistent with the inclination reported in \citet{cappellari2013}.

We find a small dark matter component to be preferred by our models. We obtain a dark matter fraction of $(7.1^{+8.1}_{-7.1})\%$ within the central (circularised) effective radius, which is lower than commonly reported for ETGs but which is consistent with the the $(15 \pm 6)\%$ previously reported for this galaxy from ATLAS\textsuperscript{3D} data. We also find our I-band M/L results to be in excellent agreement with previous ATLAS\textsuperscript{3D} spectral modelling results for this galaxy, which adds weight to the idea of there being little dark matter in this system within the observed stellar region.

We find NGC 3998 to be disc-dominated in its outer regions, with few radial orbits. We also find the orbital anisotropy to vary strongly as a function of position, with strong tangential anisotropy in the outskirts. From comparisons to simulations, we infer from both points that NGC 3998 has \textit{not} experienced significant late-time accretion of stars. We also reason from our models that some past dissipational event could have occurred. Our modelling is therefore in good agreement with the proposed FR formation path of the ATLAS\textsuperscript{3D} collaboration, in which FRs form from the quenching of high-redshift spirals.   

Overall, our results demonstrate the power of wide-field IFU instruments in studying a range of galaxy properties. We will perform a similar analysis on a wider sample in a future work.

\section*{Acknowledgements}

We wish to thank the anonymous referee for their helpful and informative report which served to greatly improve the quality of the paper. We thank Josh Adams, Guillermo Blanc and Jeremy Murphy for their help with the data reduction. We thank Kevin Luecke for his help with observing and we thank the staff of McDonald Observatory for their support. We thank Ronald Laesker for his insightful comments on comparing the SAURON and Mitchell kinematics. NFB was supported by STFC grant ST/K502339/1 during the course of this work. NFB acknowledges support from the Max Planck Institute for Astronomy in Heidelberg, Germany. AW acknowledges support from a Leverhulme Early Career Fellowship. MC acknowledges support from a Royal Society University Research Fellowship

\bibliographystyle{apj}
\bibliography{Citations}

\label{lastpage}

\end{document}